\documentclass[twocolumn,preprint]{aastex63}

\usepackage{amsmath,amstext}
\usepackage[T1]{fontenc}
\usepackage{apjfonts} 
\usepackage{natbib}
\usepackage{longtable}
\usepackage{gensymb}
\usepackage{microtype,longtable,verbatim,float,booktabs,graphicx}

\newcommand{\MgII}{MgII}

\newcommand{\Ha}{\hbox{{\rm H}$\alpha$}}

\newcommand{\HavNII}{\hbox{({\rm H}$\alpha$+{\rm [N}\kern 0.1em{\sc II}{\rm ]})}}
\newcommand{\Hb}{\hbox{{\rm H}$\beta$}}
\newcommand{\SII}{\hbox{{\rm [S}\kern 0.1em{\sc ii}{\rm ]}}}
\newcommand{\NII}{\hbox{{\rm [N}\kern 0.1em{\sc ii}{\rm ]}}}
\newcommand{\OII}{\hbox{{\rm [O}\kern 0.1em{\sc II}{\rm ]}}}
\newcommand{\OIII}{\hbox{{\rm [O}\kern 0.1em{\sc iii}{\rm ]}}}
\newcommand{\NeIII}{\hbox{{\rm [Ne}\kern 0.1em{\sc III}{\rm ]}}} \newcommand{\NeIIIvHeI}{\hbox{{\rm [Ne}\kern 0.1em{\sc III}{\rm ]}+{\rm [He}\kern 0.1em{\sc I}{\rm ]}}}
\newcommand{\NeV}{\hbox{{\rm [Ne}\kern 0.1em{\sc v}{\rm ]}}}
\newcommand{\CaVIII}{\hbox{{\rm [Ca}\kern 0.1em{\sc viii}{\rm ]}}}
\newcommand{\FeVII}{\hbox{{\rm [Fe}\kern 0.1em{\sc vii}{\rm ]}}}
\newcommand{\HeII}{\hbox{{\rm He}\kern 0.1em{\sc II}}}
\newcommand{\HeI}{\hbox{{\rm He}\kern 0.1em{\sc I}}}
\newcommand{\HII}{\hbox{{\rm H}\kern 0.1em{\sc II}}}

\newcommand{\FeX}{\hbox{{\rm [Fe}\kern 0.1em{\sc X}{\rm ]}}}

\newcommand{\NeVfull}{\NeV$\lambda$3427}
\newcommand{\CLR}{coronal line region}

\def\arrvline{\hfil\kern\arraycolsep\vline\kern-\arraycolsep\hfilneg}

\newcommand{\NeVObsLag}{$381.1^{+16}_{-22}$}
\newcommand{\NeVRestLag}{$281.7^{+12}_{-16}$}

\newcommand{\HaRestLag}{$32.5^{+12}_{-12}$}

\newcommand{\HbRestLag}{$47.3^{+29}_{-27}$}

\newcommand{\cosmo}{$\Lambda$CDM cosmology with $\Omega_{\Lambda}$ = 0.7, $\Omega_{M}$ = 0.3, and $H_{0}$ = 70~km~s$^{-1}$~Mpc$^{-1}$}

\newcommand{\PyROA}{\texttt{PyROA}}
\newcommand{\specutils}{\texttt{specutils}}

\newcommand{\JAV}{\texttt{JAVELIN}}

\newcommand*{\change}[1]{\textbf{\textcolor{blue}{[Changed: ]}}}

\shorttitle{Coronal Line Region Light Echoes}
\shortauthors{Smith et al.}

\begin{document}

\turnoffedit1

\title{\large \bf The SDSS-V Black Hole Mapper Reverberation Mapping Project: Light Echoes of the Coronal Line Region in a Luminous Quasar}

\correspondingauthor{Theodore Smith}
\email{theodore.smith@uconn.edu}

\author[0009-0003-8591-0061]{Theodore B. Smith}
\affil{Department of Physics, 196A Auditorium Road, Unit 3046, University of Connecticut, Storrs, CT 06269, USA}

\author[0000-0001-8032-2971]{Logan B. Fries}
\affil{Department of Physics, 196A Auditorium Road, Unit 3046, University of Connecticut, Storrs, CT 06269, USA}

\author[0000-0002-1410-0470]{Jonathan R. Trump}
\affil{Department of Physics, 196A Auditorium Road, Unit 3046, University of Connecticut, Storrs, CT 06269, USA}

\author[0000-0001-9920-6057]{Catherine~J.~Grier}
\affiliation{Department of Astronomy, University of Wisconsin-Madison, Madison, WI 53706, USA} 

\author[0000-0003-1659-7035]{Yue Shen}
\affiliation{Department of Astronomy, University of Illinois at Urbana-Champaign, Urbana, IL 61801, USA}
\affiliation{National Center for Supercomputing Applications, University of Illinois at Urbana-Champaign, Urbana, IL 61801, USA}

\author[0000-0002-6404-9562]{Scott F. Anderson}
\affiliation{Astronomy Department, University of Washington, Box 351580, Seattle, WA 98195, USA}

\author[0000-0002-0167-2453]{W. N. Brandt}
\affiliation{Department of Astronomy \& Astrophysics, 525 Davey Lab, The Pennsylvania State University, University Park, PA 16802, USA}
\affiliation{Institute for Gravitation and the Cosmos, The Pennsylvania State University, University Park, PA 16802, USA}
\affiliation{Department of Physics, 104 Davey Lab, The Pennsylvania State University, University Park, PA 16802, USA}

\author[0000-0001-9776-9227]{Megan C. Davis}
\affil{Department of Physics, 196A Auditorium Road, Unit 3046, University of Connecticut, Storrs, CT 06269, USA}

\author[0000-0002-4459-9233]{Tom Dwelly}
\affiliation{Max-Planck-Institut f{\"u}r extraterrestrische Physik, Giessenbachstra\ss{}e, 85748 Garching, Germany}

\author[0000-0002-1763-5825]{P. B. Hall}
\affiliation{Department of Physics and Astronomy, York University, Toronto, ON M3J 1P3, Canada}

\author[0000-0003-1728-0304]{Keith Horne}
\affiliation{SUPA Physics and Astronomy, University of St. Andrews, Fife, KY16 9SS, Scotland, UK}

\author[0000-0002-0957-7151]{Y. Homayouni}
\affiliation{Department of Astronomy \& Astrophysics, 525 Davey Lab, The Pennsylvania State University, University Park, PA 16802, USA}
\affiliation{Institute for Gravitation and the Cosmos, The Pennsylvania State University, University Park, PA 16802, USA}

\author[0000-0002-0913-3729]{J. McKaig}
\affiliation{X-ray Astrophysics Laboratory, NASA Goddard Space Flight Center, Code 662, Greenbelt, MD 20771, USA}
\affiliation{Oak Ridge Associated Universities, NASA NPP Program, Oak Ridge, TN 37831, USA}

\author[0000-0002-6770-2627]{Sean Morrison}
\affiliation{Department of Astronomy, University of Illinois at Urbana-Champaign, Urbana, IL 61801, USA}

\author[0000-0001-9616-1789]{Hugh W. Sharp}
\affiliation{Department of Physics, 196A Auditorium Road, Unit 3046, University of Connecticut, Storrs, CT 06269, USA}

\author[0000-0002-9508-3667]{Roberto J. Assef}
\affiliation{Instituto de Estudios Astrof\'isicos, Facultad de Ingeniera\'ia y Ciencias, Universidad Diego Portales, Av. Ej\'ercito Libertador 441, Santiago, Chile 8370191}

\author[0000-0002-8686-8737]{Franz E. Bauer}
\affiliation{Instituto de Alta Investigaci{\'{o}}n, Universidad de Tarapac{\'{a}}, Casilla 7D, Arica, Chile}

\author[0000-0002-6610-2048]{Anton M. Koekemoer}
\affiliation{Space Telescope Science Institute, 3700 San Martin Dr., Baltimore, MD 21218, USA}

\author[0000-0001-7240-7449]{Donald P.\ Schneider} 
\affiliation{Department of Astronomy \& Astrophysics, 525 Davey Lab, The Pennsylvania State University, University Park, PA 16802, USA}
\affiliation{Institute for Gravitation and the Cosmos, The Pennsylvania State University, University Park, PA 16802, USA}

\author[0000-0002-3683-7297]{Benny Trakhtenbrot}
\affiliation{School of Physics and Astronomy, Tel Aviv University, Tel Aviv 69978, Israel}

\author[0000-0002-9790-6313]{Hector Javier Ibarra-Medel}
\affiliation{Instituto de Astronomıa, Universidad Nacional Aut\'{o}noma de M\'{e}xico, A.P. 70-264, 04510, Mexico, D.F., M\'{e}xico}

\author[0000-0002-1656-827X]{Castalia Alenka Negrete Pe\~{n}aloza}
\affiliation{Instituto de Astronomıa, Universidad Nacional Aut\'{o}noma de M\'{e}xico, A.P. 70-264, 04510, Mexico, D.F., M\'{e}xico}

\begin{abstract}
We present a reverberation mapping analysis of the coronal line \NeVfull\ emitting region of the quasar COS168 (SDSS J095910.30+020732.2). \NeVfull\ is known as one of the "coronal lines," which are a species of emission lines present in AGN spectra with high ionization potentials ($\geq$~100 eV) that can serve as tracers for AGN activity. The spatial extent of the \CLR\ has been studied with only spatial resolving techniques that are not sensitive to the innermost regions of AGN. Through our reverberation mapping analysis of \NeV$\lambda$3427, we measure a nominal `optimal emission radius' for \NeV$\lambda$3427 of \NeVObsLag\ light days (observed-frame). We place the \CLR\ in context with other AGN regions by comparing it with the characteristic radius of \Ha, the dust-sublimation radius, and the dusty torus. The \CLR\ is located at a larger radius from the black hole than the characteristic radius of the dusty torus, measured using a torus-radius luminosity relationship. The virial product ($v^2 R/G$) of both \Ha\ and \NeV$\lambda$3427 is consistent within the uncertainties, implying that the \CLR, as probed by the \NeV$\lambda$3427 line, may be in a virialized orbit that is dominated by the gravitational potential of the black hole. This plausibly suggests that coronal lines could be an effective method for estimating black hole masses.

\

\textit{Resubmitted to ApJ after responding to referee comments}

\end{abstract}



\section{Introduction}

The majority of massive galaxies harbor a supermassive black hole (SMBH) at their centers \citep{Magorrian1998, Kormendy2013}. Central supermassive black holes that actively accrete matter are known as active galactic nuclei (AGN). This process of accretion releases enormous amounts of energy and makes AGN some of the most powerful sources of electromagnetic radiation in our universe. This extraordinarily powerful radiation shines upon the gas surrounding the black hole and gives rise to the common broad emission lines that are characteristic of AGN \citep{Seyfert1943}. AGN have strong UV and X-ray emission that far exceeds the ionizing radiation generated by stars and other sources. In addition to broad emission lines, the existence of exotic "coronal lines" with high ionization potentials is often thought of as a hallmark sign of an AGN (e.g. \citealt{Negus2023, Satyapal2023}).

\begin{figure*}[t]
\epsscale{1.15}
\plotone{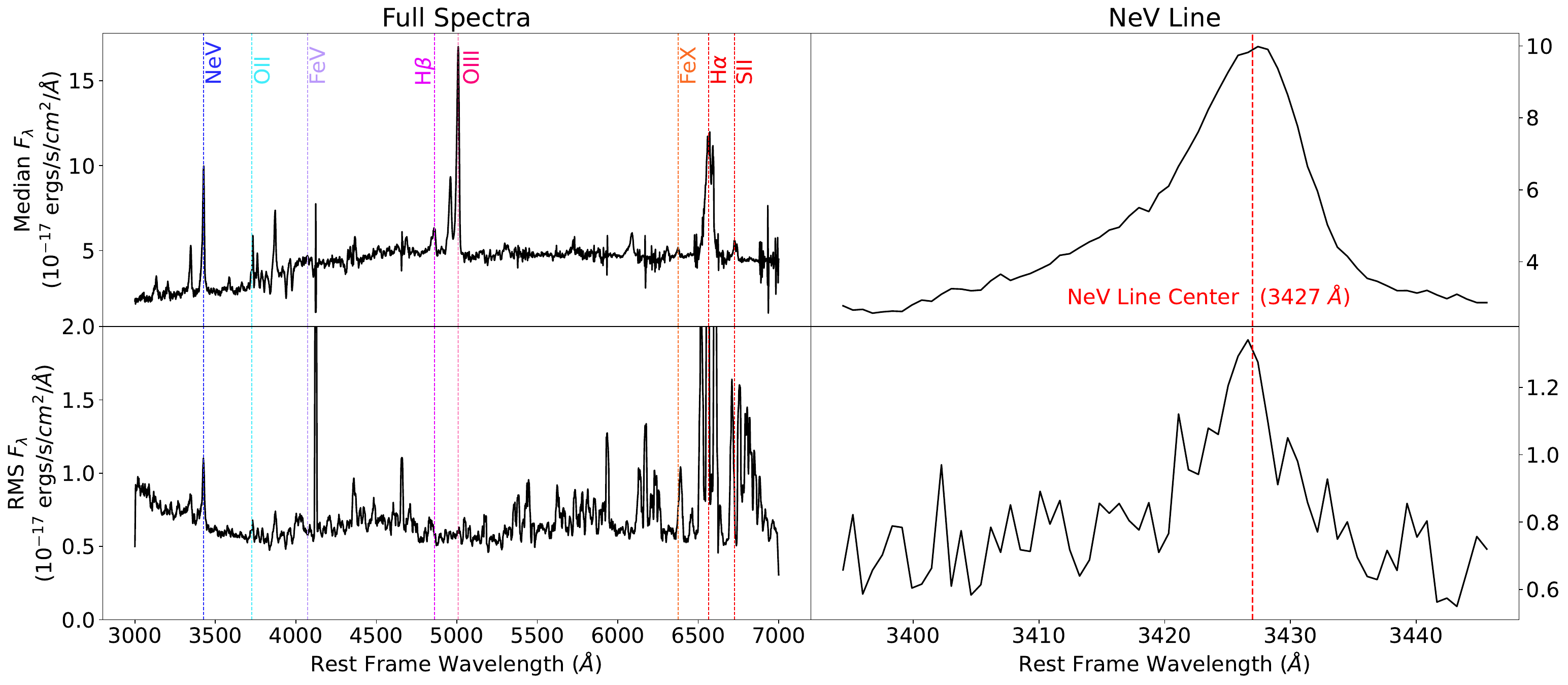}
\figcaption{Median (top) and RMS flux (bottom) spectra for COS168. The left panels show the full wavelength range of the SDSS-V spectra and the right panels show the region around the \NeVfull\ line. The full RMS spectrum is boxcar-smoothed by 13 pixels.\label{Fig1}}
\end{figure*}



Coronal lines are a species of emission lines with ionization potentials $\geq$ 100 eV. The name comes from the fact that the emission lines were first seen in the hot corona of the sun \citep{Osterbrock1989}. \edit1{Coronal lines have been shown to be rare in observed galaxies, with only around 4.5\% of galaxies in SDSS containing one measurable coronal line \citep{Reefe2022, Reefe2023, Doan2025}}. Due to their high ionization potentials, coronal lines represent a good tracer of AGN \citep{Erkens1997, Mazzalay2013, Molina2021, Reefe2022, Negus2023}. For example, \cite{Negus2023} reported that out of galaxies that exhibit \NeV$\lambda$3427 in their spectra, 87.5\% host AGN. \edit1{Unlike the standard broad lines such as \Ha, coronal lines like \NeVfull\ are a more robust marker of AGN as they cannot be produced in star formation. \NeVfull\ in particular, can be used for studies in obscured AGN as neon does not deplete onto dust grains and can be visible in spectra where broad lines are absent.}


Based on photoionization arguments, the coronal-line region has been thought to exist somewhere between the broad-line region (BLR) and the far-extended narrow-line region (NLR); \citep{Riffel2006,Baskin2014,McKaig2024}. Past estimates through, integral field unit (IFU) spectroscopy, of where coronal lines are emitted range from 3 pc to up to 250 pc \citep{Mazzalay2013}. \citet{Negus2021} used integral field unit spectroscopy from the Sloan Digital Sky Survey (SDSS) MaNGA project \citep{Bundy2015} to demonstrate that the coronal line region has an average distance of 6.6~kpc from the central black hole for a typical AGN luminosity of $L_{bol} \sim5\times  10^{44} \rm \ erg\ s^{-1}$. In later work with the MaNGA survey, \cite{Negus2023} reported that the average size of the \CLR\ for \NeV$\lambda$3427 is 1.9 kpc, although for many galaxies in both \cite{Negus2021} and \cite{Negus2023}, much of the coronal line emission is spatially unresolved in the IFU data, thus the measurement may represent a maximum radius rather than an average location. However, photoionization modeling has shown that the line-luminosity for \NeV$\lambda$3427 rises as a function of radius to a peak around $\sim$~0.4pc \citep{McKaig2024}. Additionally, \cite{Gravity2021} examined the high-ionization coronal \CaVIII\ line in NCG~3783 and measured a characteristic size of $\sim$~0.4 pc, placing the \CLR\ between the BLR and the NLR. \edit1{Further studies have placed the \CLR\ anywhere from on the scales of pc to kpc \citep{Mazzalay2010, Muller-Sanchez2011, Muller-Sanchez2018}}. The wide range of uncertainty in these previous measurements remains to be resolved, which motivates the performance of this reverberation mapping work on coronal line emission.

The inner environment of AGN is challenging to resolve spatially, with very few exceptions, such as M87 \citep{EHT2019}. Instead of spatially resolving the inner environments of AGN, we can trade spatial resolution for temporal resolution with reverberation mapping (RM; e.g., \citealp{Blandford1982, Peterson1993, Cackett2021}). Reverberation mapping utilizes the intrinsic fluctuations in the continuum and the fact that these fluctuations drive similar fluctuations in the emission from regions located some distance away. \edit1{Coronal lines have been shown to have such variability on measurable timescales, further motivating the use of RM \citep{Veilleux1988, Landt2015b}}. RM has traditionally been used to measure the size, location, and structure of the BLR and to determine the mass of black holes (e.g. \citealt{Bentz2010, Du2014, Grier2017a, Shen2024}). The broad emission lines fluctuate in response to the variations in the continuum after some time delay $\tau$, which is the light travel time from the central engine to the BLR. Using the lag, $\tau$, along with the speed of light, allows a measurement of the ``optimal emission radius'' of a particular emission species.




The use of RM is not limited to measuring the size of the BLR; it can also be used to measure the size of any emission species that fluctuates in response to the continuum on timescales less than that of the monitoring period. We do not know the radius of the \CLR; it is only measurable by RM if it has a size with a light-crossing time that is less than the monitoring duration. Essentially, a \CLR\ with a size $\simeq$ kpc as implied by IFU observations \citep{Negus2021, Negus2023} would be impossible to measure with RM. However, if the \CLR\ has a size on the order of pc or less, as implied by photoionization modeling \citep{McKaig2024}, RM of the \CLR\ is possible.

This paper presents RM results of the \NeV$\lambda$3427 coronal line of the luminous quasar, SDSS J095910.30+020732.2 (hereafter COS168). Section~\ref{Sec2} describes the spectroscopy, photometry, and spectral decomposition of COS168. Section~\ref{Sec3} presents variability measurements of the \NeVfull\ light curve, our lag measurement methodology, and results. 
Section~\ref{Sec4} demonstrates that the measured size of the \NeVfull\ emitting region is beyond the BLR and the emissivity-weighted radius of the torus, and demonstrates that the \NeVfull\ emitting region appears to be virialized. Lastly, Section~\ref{Sec5} summarizes our results.

Throughout this work, we assume a \cosmo.


\section{Data}
\label{Sec2}

\subsection{BHM-RM Spectroscopy}
\label{sec:BHMRM spec}

We use spectroscopic data from the ongoing Black Hole Mapper Reverberation Mapping (BHM-RM) Project (Trump et al. in prep), which is a part of the fifth generation \citep{Kollmeier2025} of the Sloan Digital Sky Survey (SDSS-V, \citealp{York2000}). Until 2021, the spectroscopic data for BHM-RM were taken at the Apache Point Observatory using the plate-based, fiber-fed, BOSS spectrograph \citep{Smee2013} on the 2.5m telescope \citep{Gunn2006}. After 2021, the data were taken with the new SDSS-V robotic focal plane system \citep{Sayres2022}. The data were reduced with v6\_2\_0 of IDLspec2d (\citealp{Bolton2012, Dawson2013}; Morrison et al. in prep). The BHM-RM spectra range from $\sim$3800 to $\sim$10200 \AA\ with a spectral resolution ranging from R $\sim$ 1500 in the blue and R $\sim$ 2500 in the red end of the spectrum. The data for COS168, in this study, span four years, from 2021-2024, with a total of 78 epochs of observation. The average monitoring cadence within the seasonal availability for this study was $\sim$ 6.3 days.
~~
~~

\subsection{Properties of COS168}

COS168 is a luminous ($L_{\rm {bol}} = 3.71 \times 10^{45}$ erg s$^{-1}$) quasar located at R.A. = $9^{\rm h} 59^{\rm m} 10^{\rm s}.31$ and Dec = $2^{\circ}7'32.24''$ (J2000) with a spectroscopic redshift of $z$ = 0.3528. COS168 exhibits prominent coronal line emission in the \NeV$\lambda$3427 line in the median and RMS spectra. COS168 was previously noted to have prominent \NeV\ and \FeVII\ emission lines by \citet{Lanzuisi2015}. We chose to study COS168 since it exhibits strong \NeVfull\ variability with respect to \OIII. Figure~\ref{Fig1} shows the median and RMS profiles of the \NeV$\lambda$3427 line as well as the median and RMS spectra across the entire wavelength range. Qualitatively, the \NeV$\lambda$3427 line has strong peaks in both the median and RMS spectra, indicating that it varies on the timescales we can study with.

\edit1{COS168 also exhibits weak emission from \FeVII$\lambda\lambda3760,6086$ (median SNR $\sim 5$ at each epoch) however they are much weaker than \NeVfull\ (median SNR $\sim 30$) and are not present in the RMS spectra. This is expected as it has been shown that on average, \NeVfull\ is the brightest and most common of the coronal lines \citep{Vanden2001, Reefe2022, McKaig2024, Doan2025}. Given their weakness, we choose not to examine the \FeVII$\lambda\lambda3760,6086$ lines in this study.}

\edit1{For the purposes of fitting the \NeVfull\ line, it is important to examine the line profile. Coronal lines have been shown to exhibit blue-ward asymmetries in their line profiles, indicative of outflows  \citep{Erkens1997, Rodriguez2006, Rodriguez2011, Kynoch2022, Doan2025, Matzko2025}. The \NeVfull\ line profile in COS168 does exhibit weakly increased emission in the blue wing of its spectra; however, that emission is sub-dominant (i.e., below half-max) and does not significantly affect the FWHM measurement. Thus, going forward, we opt for simplicity and choose to fit the \NeVfull\ line with a single Gaussian.} 



\subsection{Spectral Decomposition}
\label{Sec:spectral_decomp}

To isolate the \NeV$\lambda$3427 line, we define a window for spectral decomposition using the composite SDSS quasar spectrum (see Table 2 in \citealt{Vanden2001}). Along with \NeVfull, we define windows for spectral decomposition for \Hb\, \Ha\, \OII, and \OIII\ as shown in Table \ref{Tab:spectral_decomp}, to isolate the continuum. We then subtract the spectroscopic continuum flux from the spectrum by fitting a first-order polynomial to the spectrum based on the median of the continuum over 10 \AA\ from the line-free regions blueward and redward of each line. We then fit the isolated \NeV$\lambda$3427 line with a one-dimensional Gaussian model. As a note, we do not see convincing evidence of a complex \NeVfull\ line profile, consisting of narrow and broad components.

\begin{figure*}[t]
\plotone{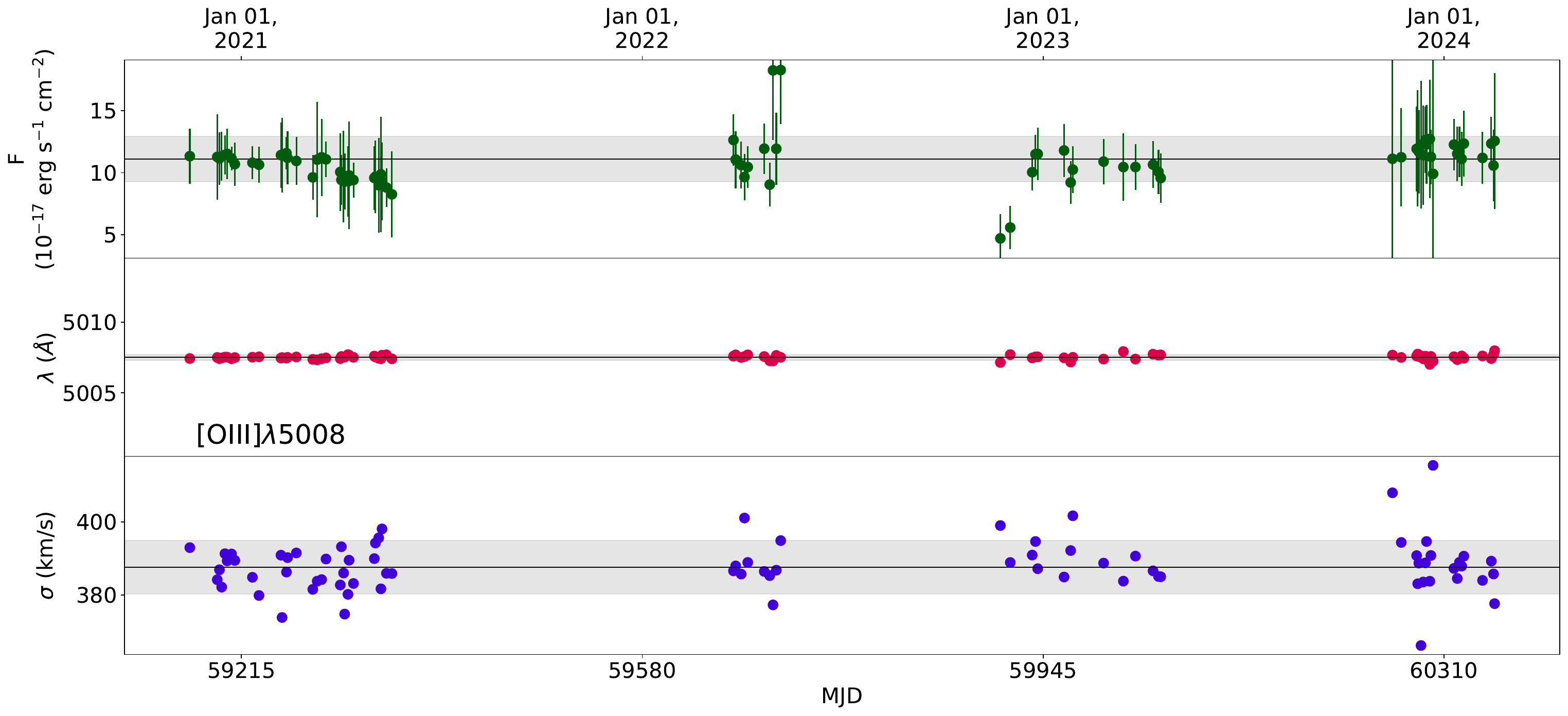}
\epsscale{1.1}
\figcaption{Gaussian fit parameters of the \OIII$\lambda$5007 emission line in COS168 (line flux, line center, and line width) as well as the median over all epochs for each parameter (black line) and the normalized median absolute deviation (NMAD; grey areas). The line center and line width are stable so we use line flux as a flux calibrator.\label{fig:oiii_params}}
\end{figure*}

\begin{table}[h]
\begin{tabular}{lcc}

\toprule
Emission Line &  Lower Bound (\AA)&  Upper Bound (\AA) \\
\midrule
 \NeV$\lambda$3427 & 3394 &  3446  \\
  \Hb & 4760  &   4980  \\
  \Ha &  6400 &   6765 \\
  \OIII & 4982 & 5035 \\
  \OII  & 3714  & 3740 \\
\bottomrule
\end{tabular}

\caption{Spectral decomposition window for the emission lines studied in this project.}
\label{Tab:spectral_decomp}

\end{table}

\subsection{Calibrations Based on Narrow Emission Lines}

SDSS-V quasars have been shown to have epoch-dependent changes in the spectrophotometric calibration \citep{Shen2015, Fries2023}. We therefore investigated the stability of the \OIII$\lambda$5007 line for COS168 under the assumption that \OIII\ is constant over short timescales \citep{Foltz1981, Peterson1982}. We chose to fit a single Gaussian to the \OIII$\lambda$5007 line with the same fitting procedure described in Section \ref{Sec:spectral_decomp}. Figure \ref{fig:oiii_params} presents the line profile parameters of the \OIII$\lambda$5007 line over the course of our monitoring period. The line center and line width are stable, so calibration of these parameters is not necessary for this object \citep{Fries2023}. The flux of \OIII$\lambda$5007 is not constant throughout the monitoring period, despite being emitted from a relatively extended region in this AGN. The apparent \OIII\ flux variability is due to spectrophotometric issues rather than real variability, as supported by observations of NGC 5548 \citep{Peterson2013}, which shows that \OIII\ varies on timescales much larger than the timescales studied in this work. Additionally, since COS168 is much more luminous than NGC 5548, it is likely that its \OIII\ emission region is larger \citep{Hainline2013}. We therefore calibrate our spectra based on the assumption of a constant \OIII$\lambda$5007 line, which is often used as a flux calibrator \citep{U2022, Fries2023, Shen2015}.

We perform second-order polynomial corrections to the spectra to require that the \OIII$\lambda$5007 line is constant over the monitoring period, following a similar procedure to \cite{Fries2023}. Briefly, we multiply the flux at every wavelength over every epoch by a corrective factor that is determined by the \OIII$\lambda$5007 flux at that epoch. The correction factor for flux at each epoch $i$ is the median flux of the \OIII$\lambda$5007 line divided by the flux of the \OIII$\lambda$5007 line at that epoch ($\overline{f}\OIII/f\OIII_{i}$). We also calibrate the wavelength of the spectra using the difference between the median \OIII$\lambda 5007$ line center (see Figure \ref{fig:oiii_params} center panel). 


\subsection{ZTF Photometry}
\label{sec:ZTF}
We used the $g$ and $r$ time-domain photometry observed by the Zwicky Transient Facility (ZTF; \citealp{Masci2019}) to model the continuum as our driving light curve. The data cover a time span from 2019 to 2024 and were obtained from the 22$^{\rm nd}$ data release from ZTF\footnote{https://www.ztf.caltech.edu/ztf-public-releases.html}. To merge the two ZTF photometric light curves into a single curve, we take the average flux density in the 2020 monitoring period for both the $g$ and $r$-band light curves and create a scale factor, which is the average $r$-band flux density over the average $g$-band flux density. We then scale the $g$-band light curve by multiplying it by this scale factor, resulting in a single photometric light curve consisting of the individual $g$-band and $r$-band light curves.

\section{Analysis}
\label{Sec3}

\subsection{Variability of \NeVfull\ Line}
We first quantify the variability of the \NeVfull\ emission line in COS168 by calculating the fractional variability of the \NeVfull\ emission line. The fractional variability is defined as:
\begin{equation}
     \% \ \mathrm{Variable} \equiv \frac{\sqrt{\mathrm{NMAD}(f)^2 - \mathrm{med}(\sigma)^2}}{\mathrm{med}(f)^2}
\end{equation}
\noindent where NMAD is the normalized median absolute deviation, $f$ is the observed flux measurement, and $\sigma$ is the uncertainty in each flux measurement. For this measurement and others to come, we derive uncertainties in \NeVfull\ flux measurements from a Monte Carlo simulation of the \NeVfull\ line. Following this, we also calculate a metric of how well variability is measured, known as SNR2 \citep{Shen2019}, defined as: SNR2 $\rm =\sqrt{\chi^2 - dof}$, where $ \chi^2 \equiv \sum_i (f_i - \langle f_i \rangle)^2/\sigma_i^2$ and dof corresponds to the degrees of freedom equal to $N -1$. We find that the \NeVfull\ line in COS168 has a fractional variability of $\sim$8\% and an $\rm SNR2 \approx 29$, which is greater than the recommended threshold value ($\rm SNR2 \geq 20$) for a significant detection \citep{Shen2019}. \edit1{Additionally, we measure the fractional variability of \Ha\ to be $\sim 7.3\%$ with an SNR2 $\approx 12$, showing that for this object, the variability of \NeVfull\ is comparable to that of known broad lines like \Ha.}

The \NeVfull\ light curve (see Figure \ref{fig:PyROA fits} top panel) exhibits apparent variability on timescales far shorter than that of the baseline timescale. Paired with the fractional variability and SNR2 measurements, suggests that the \NeVfull\ emission region in COS168 may be responding to continuum variability on similar timescales and might be located closer to the center of the AGN than the farther extended, less variable, NLR.

Interestingly, we find that the fractional continuum variability measured at $5100$ \AA\ in COS168 is about half that of \NeVfull\ (4\% fractional variability, SNR2$\ \approx 11$), implying that \NeVfull\ is more variable than the optical continuum. This suggests that \NeVfull\ might be responding to UV or far UV continuum variability. This is somewhat expected given the immense energy required to produce \NeVfull\ emission \edit1{as well as the increased variable flux in the blue end of the RMS spectra}.

\subsection{Lag Measurement Methodology}
 
We use the ZTF photometric light curve for the driving light curve and the BHM-RM spectroscopic light curves as responding light curves in our RM analysis. We use \PyROA\footnote{https://github.com/FergusDonnan/PyROA} \citep{Fergus2021} to fit our light curves to measure emission-line lags.



\PyROA\ uses a Bayesian Markov chain Monte Carlo (MCMC) approach to model quasar variability with a running optimal average (ROA) function. The ROA is applied within a Gaussian window function, where the width is a prior set by the user. This ROA window function highly weights points near the point of interest, while neglecting to consider points far from the point of interest. \PyROA\ requires the user to input upper and lower limits for uniform priors on five parameters: $A$ (rms flux of the light curves), $B$ (mean flux of the light curves), $\tau$ (the observed time delay range of the light curves being modeled), $\Delta$ (the range of acceptable window function width values), and $\sigma$ (an extra error rescaling parameter). As the \CLR\ is thought to be more extended than the BLR, we increase the upper limit of $\tau$ to 1000 days when measuring \NeV, while setting the upper limit of $\tau$ to 250 days for \Ha\ and \Hb. Our \PyROA\ priors are shown in Table \ref{tab:pyroa}. Previous work has established that \PyROA\ lag measurements (\citealp{Fergus2021, Fries2023}; Sharp et al. in prep) are consistent with lag measurements produced by other methodologies such as \JAV\ \citep{Zu2010} and ICCF \citep{Peterson2004, Sun2018}.

\begin{table}[h]
\label{tab:pyroa}
\centering

\begin{tabular}{cc}
\toprule
\PyROA~Parameter & Value \\
\midrule
A  &  [0, 2.0] \\
B  &  [0, 2.0] \\
$\tau$  &  [0, 1000] \\
$\Delta$ & [0.05, 50] \\
$\Delta_{init}$ &  10 \\
$\sigma$  &  [0, 5.0] \\

\bottomrule

\end{tabular} 
\caption{Values of different lower and upper limits to the uniform \PyROA\ priors used in our RM analysis. For \Ha\ and \Hb, the $\tau$ parameter was set to [0, 250] since the luminosity of COS168 implies a BLR of less than a few hundred light days \citep{Bentz2010} 
}
\end{table}

\subsection{Lag Measurements}
\label{Sec: 3.2 lag measurments}
\begin{figure*}[t]
    \plotone{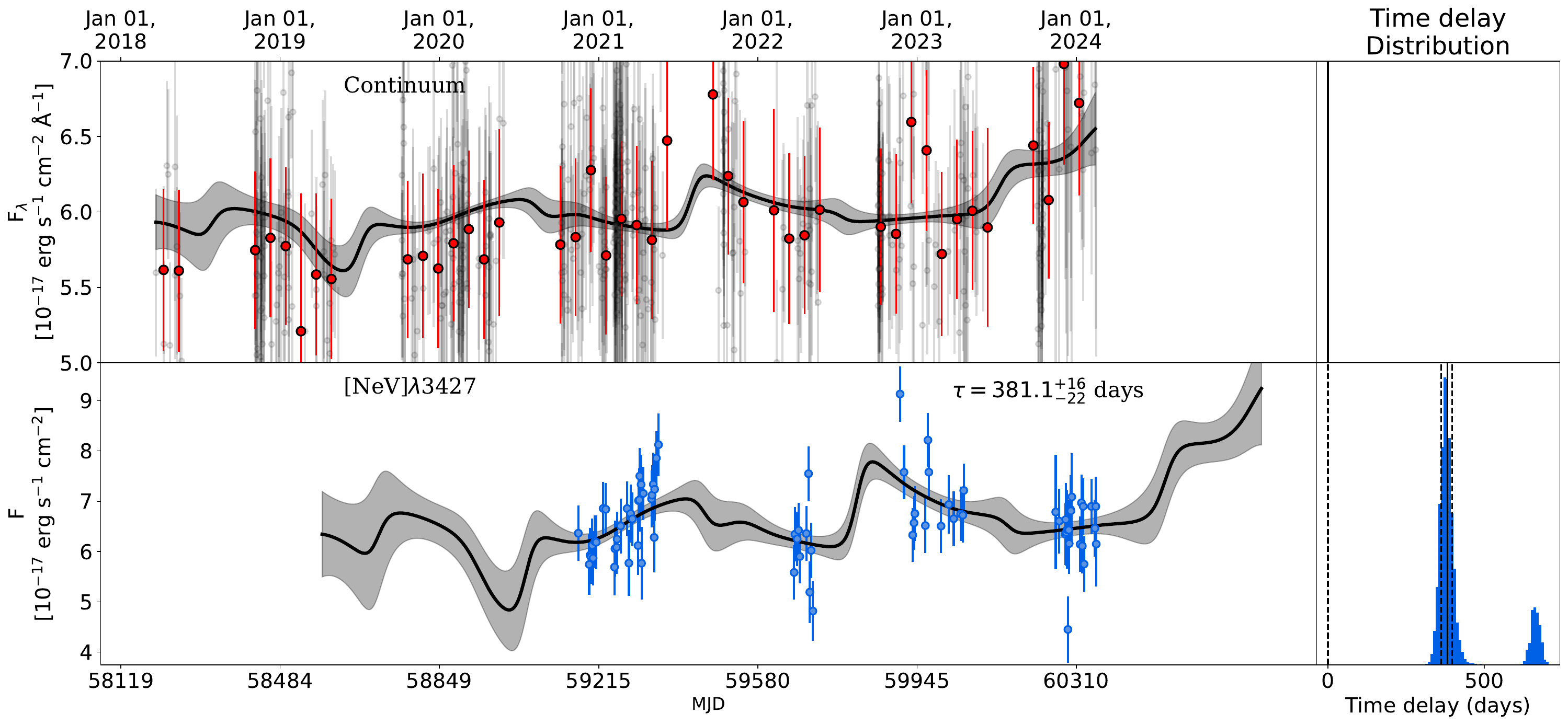}
    \plotone{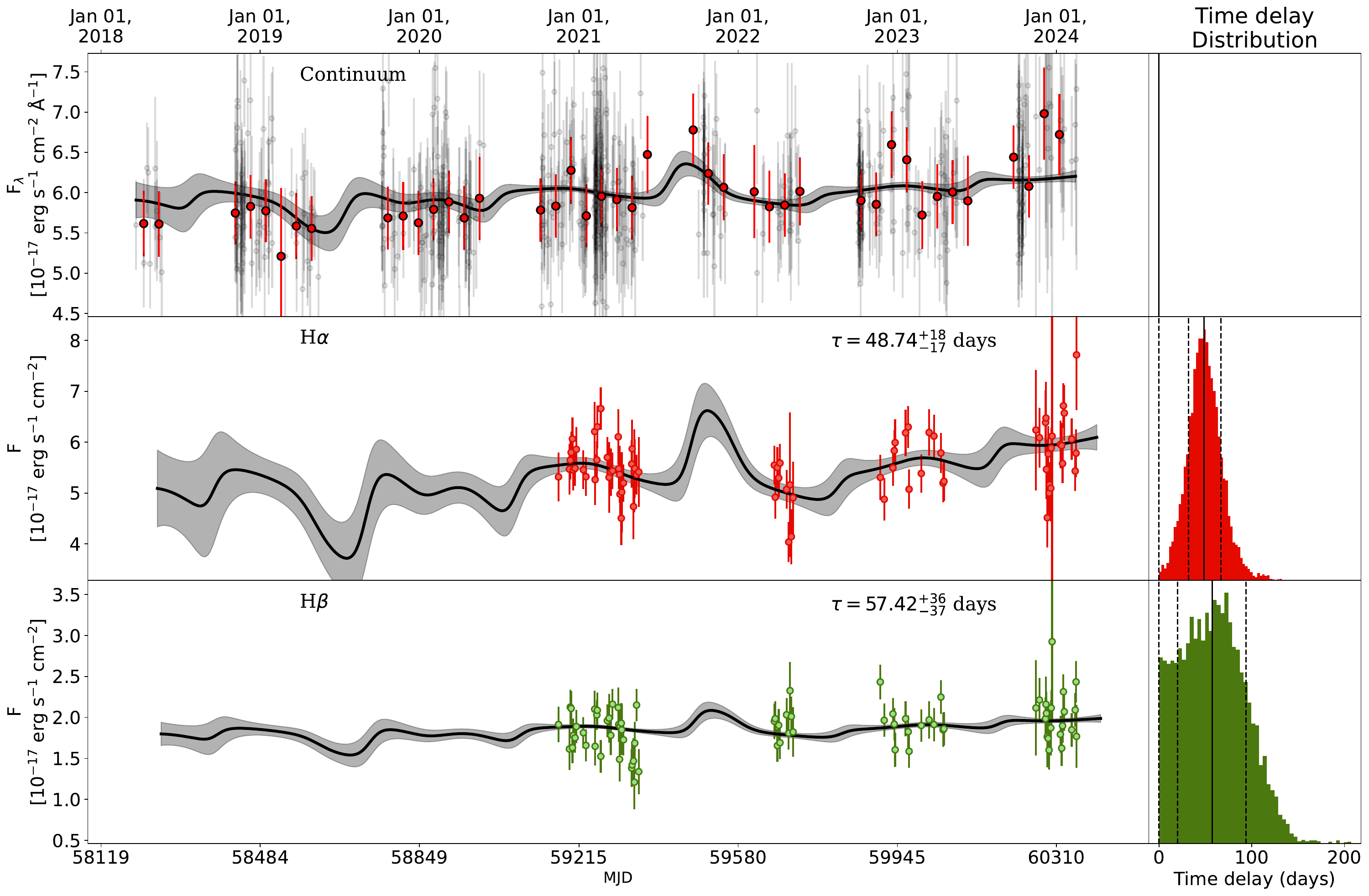}
    \epsscale{1.1}
    \figcaption{PyROA best-fit light curves for \Ha\ and \Hb\ (bottom), as well as \NeVfull\ (top). The top panel in each plot shows the continuum light curve binned every 30 epochs (with gray points representing the raw data), with the best-fit \PyROA\ model as a solid black line with a corresponding error envelope. The bottom panel in each plot shows the measured emission-line light curve with its corresponding \PyROA\ model. The \PyROA\ best-fit observed-frame time delays are shown in the emission line light curve panels with their corresponding error margins. The bottom right panel displays the posterior time delay distribution for each emission-line light curve with median and 68$^{\rm th}$ percent errors as solid and dashed lines, respectively.  \label{fig:PyROA fits}}
\end{figure*}

We measure the observed-frame time delay of \NeV$\lambda$3427 to be \NeVObsLag\ days. Figure \ref{fig:PyROA fits} displays the light curves of the \NeVfull\ emission line and the continuum along with the \PyROA, best-fit model for each light curve. Figure \ref{fig:PyROA fits} also shows the probability distributions for the best-fit observed frame lag of the \NeVfull\ emission line. The lag is well-constrained with 86\% of the probability distribution in a single peak at $\tau=381$~days. There is a weak secondary peak at $\tau=765$~days that corresponds to an alias solution that adds a year to the measured lag. The uncertainties in the lag are measured from the 16th and 84th percentiles of the lag probability distribution, excluding the lesser alias peak.

To place the \NeVfull\ emission region in context, we additionally measure the time delays of the Balmer lines. The observed frame time delays for \Ha\ and \Hb\ are $44.0^{+16.8}_{-16.7}$ and $64.0^{+39.0}_{-36.0}$ days, respectively. Figure~\ref{fig:PyROA fits} presents the best-fit \PyROA\ model for \NeV$\lambda$3427, \Hb, and \Ha, respectively. Table~\ref{tab3} lists our measured time delays for \NeV$\lambda$3427, \Ha, and \Hb. The best-fit model for \Hb\ is relatively unconstrained, plausibly given the lack of variability in the spectrum over the course of our monitoring period. While it is still constrained to be much less than the \NeVfull\ lag, this result does not represent a good measurement of the \Hb\ emission region. 

\begin{figure*}
    \plotone{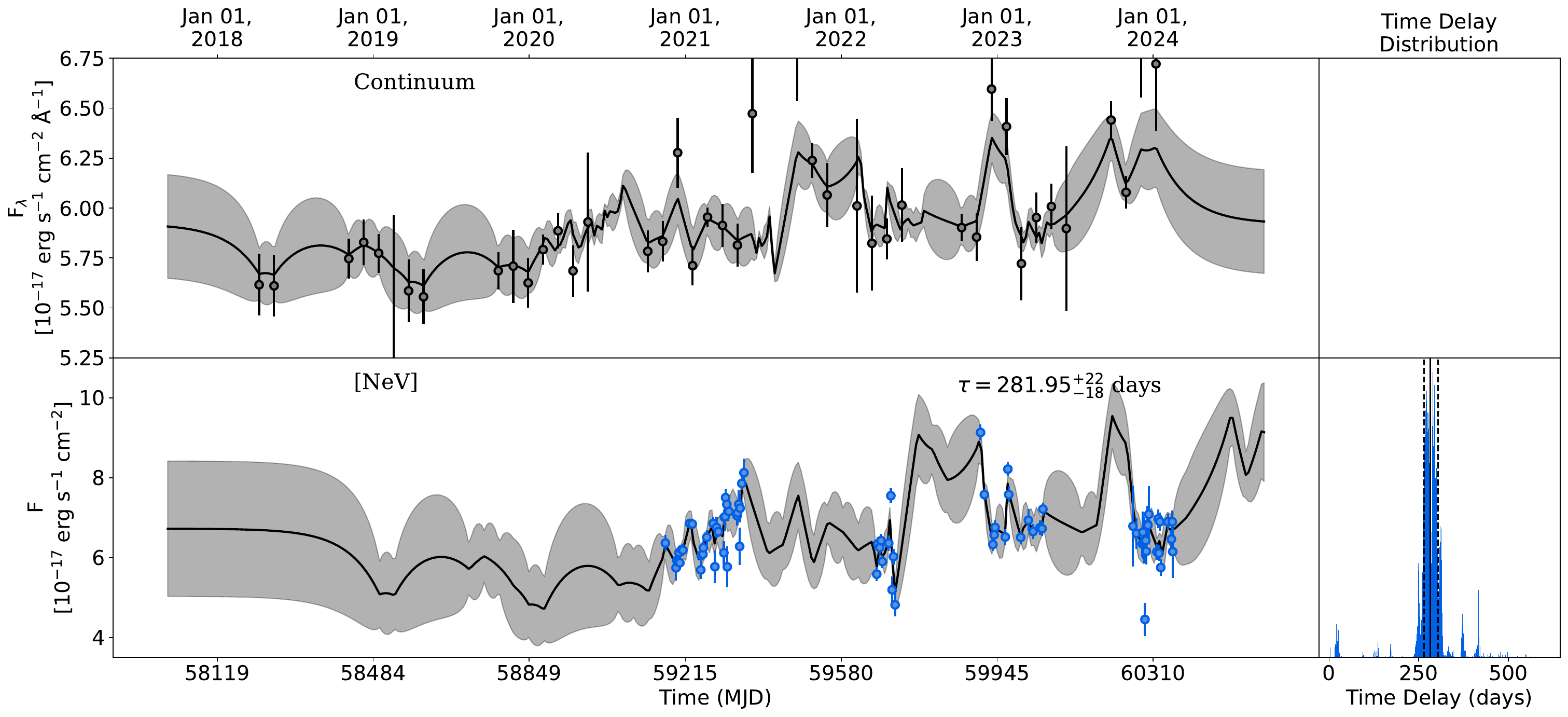}
    \epsscale{1.1}
    \figcaption{\JAV\ best fit light curves for \NeVfull. The continuum data is the same binned data from Figure \ref{fig:PyROA fits} but without the added variance from \PyROA. Much like the \PyROA\ best fit light curves, the best fit model created by \JAV\ is represented with a solid black line with a corresponding error envelope, as well as a posterior lag distribution shown in the right panels. \label{fig: JAVELIN NeV}}
\end{figure*}

\begin{table}[h]
    \centering
    \begin{tabular}{cc}
    \toprule
    Emission Line & $\tau$ (days) (rest-frame) \\
    \midrule
     \NeV$\lambda$3427    &  \NeVRestLag  \\
     \Ha     &   \HaRestLag  \\
     \Hb     &  \HbRestLag  \\
     \bottomrule
    \end{tabular}
    \caption{Rest-frame time lags for each emission line studies. The data in Figure \ref{fig:PyROA fits} demonstrate the RM done on each emission line to measure these values.}
    \label{tab3}
\end{table}

Our measured rest-frame time delay of \HaRestLag\ days for \Ha\ is consistent with the radius-luminosity relation of previous results for \Ha\ \citep{Grier2017a}. Additionally, our measured rest-frame time delay for \NeV$\lambda$3427 of \NeVRestLag\ days suggests that \NeV$\lambda$3427 is plausibly being emitted from a more extended region than the BLR.

We also measure an observed-frame lag of $282.0^{+22}_{-18}$ for \NeVfull\ using \JAV\footnote{https://github.com/nye17/javelin}, which utilizes a damped random walk (DRW) to model quasar variability and measure lags \citep{Zu2010}. The results from \JAV\ are shown in Figure \ref{fig: JAVELIN NeV}. In support of the measurement made with \PyROA, the lag results for \NeVfull\ with \JAV\ agree within $\sim\ 2\sigma$, despite their differences in lag measurement methodology.



\section{The Structure of the Coronal Line Region in COS168}
\label{Sec4}

\subsection{Comparison to Previous Measurements of Coronal Line Region Sizes}

Different studies of coronal line emission have resulted in varying numbers for the spatial extent of the \CLR; however, all previous studies have used techniques to spatially resolve the \CLR\, such as IFU spectroscopy \citep{Negus2021}, and interferometric methods \citep{Prieto2005, Gravity2021}. \cite{Negus2021} used the SDSS MaNGA IFU to spatially resolve \NeV, \FeVII, and \FeX\ emission in several AGN from 1.3-23 kpc from the center of the AGN. They find the \CLR\ to extend into the kiloparsec range, overlapping with the NLR, although most of the coronal line emission was centered in the inner kpc. \cite{Mazzalay2010} used the Hubble Space Telescope (HST) and ground-based telescopes to spatially resolve the \CLR\ and measured that the \CLR\ extends from around 10 - 230 pc. Earlier studies into coronal line emission using adaptive optics at the VLT indicated \CLR\ extension from between 20 - 230 pc \citep{Prieto2005}. \edit1{Furthermore, studies have shown evidence for a stratified \CLR\ with a variable inner emission component on pc scales and an outer emission component on the scales of 1-10 pc \citep{Gelbord2009, Kynoch2022}.}

Our measured \NeVfull\ emission region of \NeVRestLag\ days is much smaller than measurements from previous work. The aforementioned IFU studies are sensitive only to scales larger than the instrumental resolution and may represent maximal sizes for the \NeVfull\ emission. In contrast, our study is sensitive only to variable emissions associated with light-crossing times less than the monitoring duration ($\sim$1000~days), and thus is likely to probe the innermost \NeVfull\ emission regions. In general, previous IFU measurements found that most \NeVfull\ emission was unresolved \edit1{on the scales of the inner pc} of the observed AGN \citep{Mazzalay2010, Negus2021}. \edit1{Additionally, there is evidence that in a stratified \CLR, the bulk of coronal line emitting gas exists on scales $\lesssim$ 1 ly near the BLR \citep{Gelbord2009,Kynoch2022}}. These results imply that most of the coronal line emission happens on small, light-year scales, which is consistent with our measurement. Above all, our measurement indicates that most of the \textit{variable} \NeVfull\ emission happens on small (light-year) scales. Small scales are also expected to be the dominant region for coronal line emission in photoionization modeling \citep{McKaig2024} \edit1{and have been observed as the optimal emission regions for coronal lines in some galaxies \citep{Gravity2021}.}

\edit1{In a recent study done by \cite{Yin2025}, the authors performed a decade-long survey of the \FeVII\ and \FeX\ lines in a single object and detected a lag of 652 days for \FeVII. While the authors do note that a portion of the \FeVII\ flux may originate on the scale of 10 pc away, the bulk of the emission is centered in the pc range. The results presented here are consistent with our reverberation mapping results for the \NeVfull\ line in COS168.}.


\subsection{Comparison to Other AGN Structures}
\label{section:4.2}

The \NeVfull\ region is much larger than the BLR in COS168, as measured from \Ha\ (\HaRestLag\ days) and \Hb\ (\HbRestLag\ days). However, it is still unknown where our measurement of the emitting region of \NeV$\lambda$3427 lies within the structure of AGN, such as the torus and NLR. The \MgII\ region is thought to be the outer edge of the BLR and possibly truncated by the dusty torus \citep{Guo2020}. Previous studies of the \MgII\ emitting region suggest that \MgII\ is most optimally emitted from a radius 1.5-2 times that of \Ha\ \citep{Homayouni2020}. Thus, our measurement for \NeV$\lambda$3427 suggests that it may be arising from a region beyond the BLR.

\edit1{Previous studies have shown that coronal lines may be emitted near or within the dusty torus due to their obscuration in Seyfert 2 galaxies \citep{Lamperti2017}.} We compare our measured [NeV] size to the expected size of the dusty torus in COS168. We calculate the dust-sublimation radius using the following relation from \cite{Laor1993}:

\begin{equation}
\label{Eq:sublimation_radius}
    r_{\rm{min}} \simeq 0.2\left(\frac{L_{\rm bol}}{10^{46}~\rm {erg~s^{-1}}}\right)^{1/2}~\rm{pc}
\end{equation}

\noindent where $L_{\rm{bol}}$ is calculated using a bolometric correction to the optical luminosity ($L_{\rm{5100}}$) from \cite{Runnoe2012}:

\begin{equation}
\label{Eq:bol_corr}
    \rm{log_{10}}\left(\frac{L_{\rm{bol}}}{\rm{erg ~s^{-1}}}\right) = 4.89 + 0.91~\rm{log_{10}}\left(\frac{\lambda L_{\lambda}}{\rm{erg~s^{-1}}}\right)
\end{equation}
where $\lambda L_{\lambda}$ is the luminosity at 5100~\AA. For COS168, we find the dust-sublimation radius to be 143 light-days. We also calculate the characteristic radius of the torus using the torus-radius luminosity relationship from \cite{Lyu2019}:

\begin{figure*}
    \plotone{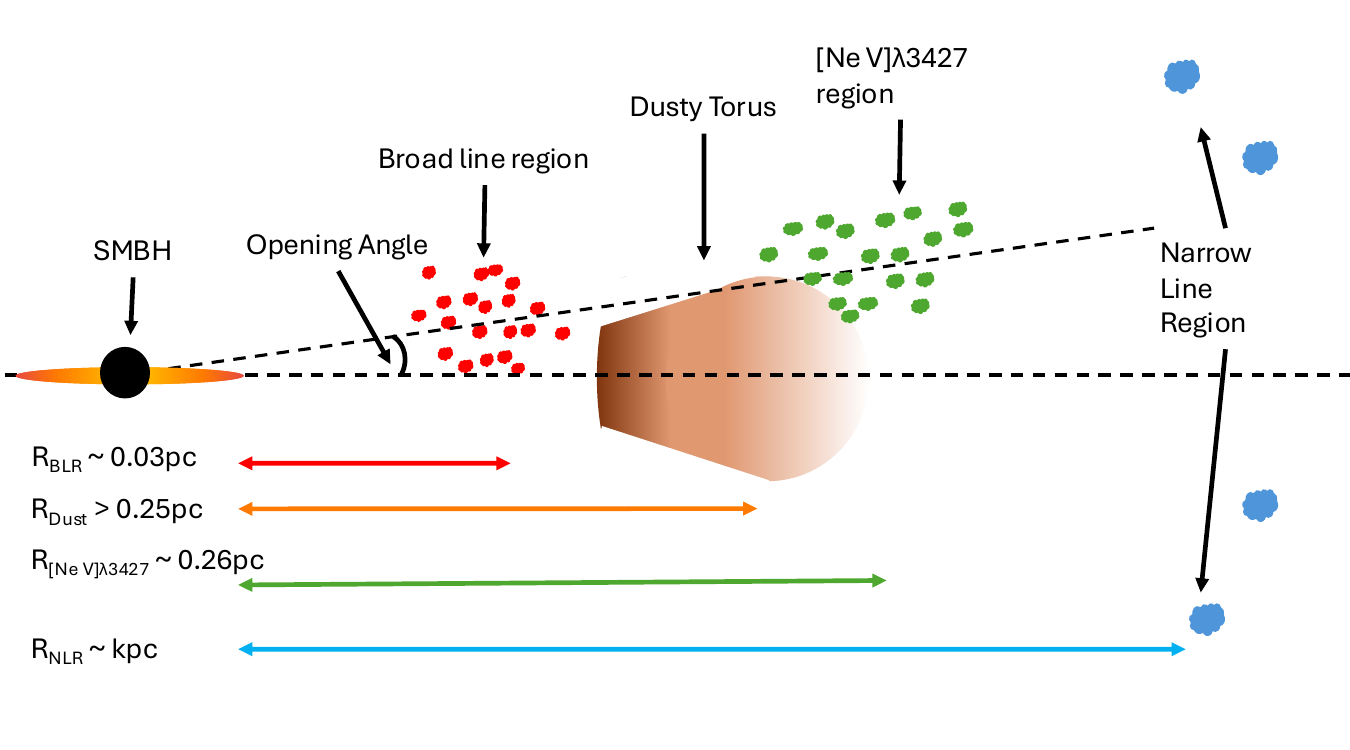}
    \figcaption{Cross-sectional diagram of the AGN COS168 (not to scale). Shown are the measured distances to the BLR from \Ha\, the dusty torus (see Section \ref{section:4.2}), and the \CLR\ from \NeV. The distance to the NLR is an estimation based on previous studies into the NLR in AGN \citep{Peterson2013}.\label{fig:cartoonfig}}
\end{figure*}

\begin{equation}
\label{Eq:torus_radius}
    \tau_{\rm{torus}}/\rm{day} = 10^{2.10 \pm 0.06} (L_{\rm{bol}}/10^{11} L_{\odot})^{0.47 \pm 0.06}
\end{equation}

where $L_{\rm{bol}}$ is calculated using Equation~\ref{Eq:bol_corr}. We find a characteristic radius of the torus to be $\sim$~297 light-days. Our measurement of the \NeV$\lambda$3427 line is somewhat larger than the dust-sublimation radius and comparable to the characteristic torus size of COS168, suggesting that the \NeV$\lambda$3427 emission is produced from beyond both the broad-line region and near or beyond the dusty torus. Figure \ref{fig:cartoonfig} depicts where the dusty torus is with respect to other regions of the AGN in COS168.

\subsection{Virialization of the \NeV$\lambda\rm{3427}$ Emitting Region}

One of the primary goals of RM is to measure the masses of AGN by probing the velocity and distance of the BLR. This only works under the assumption that the orbit of the BLR is dominated by the gravitational potential of the black hole and that we are observing a relatively uniform distribution of gas clouds across an entire orbit. Assuming the BLR is virialized, the mass of the black hole can be calculated using the virial theorem:

\begin{equation}
    \label{eq2}
    M_{BH} = f\frac{v^{2}~R_{\rm BLR}}{G}
\end{equation}
\newline
where $R_{BLR}$ is the radius of the BLR, $v$ is the width of the emission line, $G$ is the gravitational constant, and $f$ is a dimensionless factor that is introduced to characterize the unknown inclination, geometry, and kinematics. $f$ has been calibrated with spatially resolved BLR measurements \citep{Sturm2018}, AGN gas kinematics \citep{Grier2013, Woo2015}, and dynamical modeling \citep{Pancoast2014, Grier2017b}, to yield values of $f_{\sigma} = 4.47^{+1.42}_{-1.08}$ \citep{Grier2013} and $f_{\rm FWHM} = 1.12^{+0.36}_{-0.27}$ \citep{Woo2015}, \cite{Shen2024} derived a virial coefficient $f_{\sigma}$ based on all dynamical modeling AGNs at that time, which is slightly different but consistent with earlier values. Importantly, this assumption of a virialized BLR has enabled hundreds of SMBH measurements outside of the local Universe \citep{Grier2017a, Bentz2010, Du2014, Shen2024}.

\begin{figure}[h!]
    \includegraphics[width=1\linewidth]{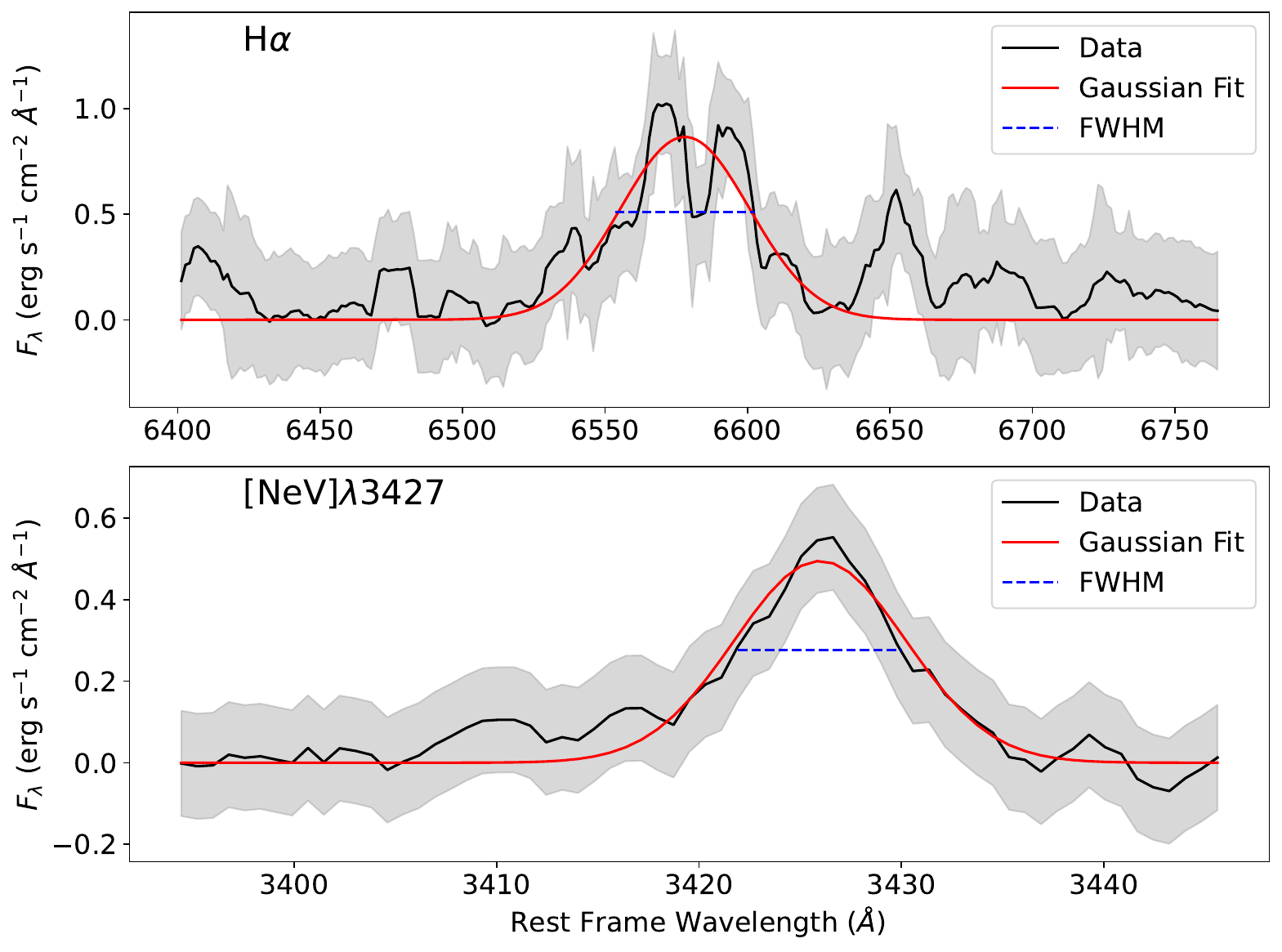}
    
    \figcaption{Line profiles of the RMS flux of the \NeVfull\ and \Ha\ line in COS168. The Gaussian fit of the emission profile and the measured FWHM of the data are shown on each plot. The \Ha\ line is particularly noisy due to sky line emission. Despite this, both the Gaussian fit and measured FWHM qualitatively fit the data.    \label{fig:nev-ha line profile}}

\end{figure}

We use the measured \NeVfull\ lag for COS~168 to similarly investigate if the \NeVfull\ line region is virialized. We expect an emission region to be virialized at small scales that are gravitationally dominated by the SMBH (as opposed to large scales, which are dominated by galaxy kinematics). We calculate the virial product ($v^2R/G$) of \NeV$\lambda$3427, and compare to \Ha, in order to probe the \CLR\ dynamics. For this comparison, we do not use \Hb\ since we do not measure a reliable \Hb\ lag for this object (see Section \ref{Sec: 3.2 lag measurments}). We use the RM measurements for $R_{\rm BLR}$ (see Table~\ref{tab3}). We measure line velocities of \Ha\ and \NeVfull\ by measuring the line width and the full width half maximum (FWHM) of each line. We measure line width by measuring the width of a Gaussian fit to an emission line in the RMS and median spectra and converting that measurement to a velocity in $\rm km\ s^{-1}$. In order to measure FWHM, we apply a median boxcar smoothing with boxcar widths of 9\AA\ for \Ha\ and 5\AA\ for \NeVfull\ to our linearized RMS spectra and leave the median spectra unchanged. We then used the FWHM routine from the \specutils\ package \citep{Earl2022} to calculate the FWHM. Measuring the FWHM of \Ha\ in particular was challenging due to the presence of highly variable sky-lines from the atmosphere present in the spectra, which can be seen in Figure \ref{fig:nev-ha line profile}. Our measured values for the virial product of the SMBH residing in COS168 are shown in Table \ref{tab4}.

\begin{table}[h]
    \centering
    \begin{tabular}{c|ccc}
     & \Ha  & RMS \NeV & Median \NeV \\
       
    \midrule
    $\sigma$ (km/s) & $1039.05^{+31.8}_{-26.0}$ & $370.778^{+2.1}_{-3.7}$  & $402.434^{+2.2}_{-3.9}$ \\
     FWHM (km/s)  & $2241.15^{+68.6}_{-56.0}$ & $771.136^{+4.0}_{-6.9}$ & $882.278^{+4.9}_{-8.5}$ \\
    V.P. ($\sigma$) &  $0.68^{+0.27}_{-0.26}$ & $0.75^{+0.09}_{-0.04}$ & $0.89^{+0.10}_{-0.05}$\\
    V.P. (FWHM) & $3.19^{+1.23}_{-1.23}$  & $2.79^{+0.30}_{-0.16}$ & $4.28^{+0.47}_{-0.24}$ \\
    
     \bottomrule
    \end{tabular}
    \caption{Measured values of the viral product using the emission lines \Ha\ and \NeV$\lambda$3427. The virial products were measured using Equation \ref{eq2} without the scale factor, $f$. Measured line width and FWHM were used for line velocity. Virial products measured using line width and FWHM are marked in the table with $\sigma$ and FWHM, respectively.}
    \label{tab4}
\end{table}



\edit1{Coronal line emission has been shown to be complex across a wide range of galaxies with outflows and off-nuclear emission (e.g., \citealp{Erkens1997, Rodriguez2006, Landt2015a}).} However, in COS168, the virial products calculated using the broad line \Ha\ and the coronal line \NeV$\lambda$3427 are \edit1{statistically consistent, implying that the \NeVfull\ line profile is consistent with virial motion.} Based on these results, it is plausible that the coronal line region is gravitationally dominated. 

\edit1{This statement is helped by the assumption that photoionization is the driving force behind \NeVfull\ emission. It has been shown that coronal line emission can be powered by shocks and outflows as opposed to AGN photoionization in AGN with spatially extended coronal line emission \citep{Pier1995, Muller-Sanchez2011, Mazzalay2013, Landt2015a, Landt2015b}. Shocks are important as a mechanism of dust destruction, which would increase the line emission of different coronal lines; however, as neon is a noble gas and thus does not deplete onto dust grains, this would not affect the emission of \NeVfull. Additionally, we acknowledge that the possibility that \NeVfull\ emission is possibly affected by AGN shocks. If AGN shocks contribute to the emission of \NeVfull, it would affect the lag measurement and consequently the virial product calculation. However, in the modeling work presented by \cite{McKaig2024}, it is shown that if strong AGN luminosity is present, \NeVfull\ emission will be dominated by photoionization with only a weak potential contribution from shocks.}


The consistency between the RMS \NeVfull\ and \Ha\ virial products implies that \NeVfull\ possibly could be used as a BH mass indicator (since $M_{B H} = f \times$ V.P.). The viral product agreement also implies that both the \NeVfull\ emission region and the BLR could plausibly have a consistent virial factor, $f$, within uncertainty, and thus have similar geometry and inclination. Additionally, the agreement between the V.P.s from the median \NeVfull\ flux profile and RMS \Ha\ suggests that our measurement shows the dominant \NeVfull\ emission source as opposed to the dominant variable emission source. Agreement from the median flux profile of \NeVfull\ also suggests that \NeVfull\ could be used for single-epoch BH mass measurements \citep{Vestergaard2006}.

\section{Conclusions}
\label{Sec5}
We have provided measurements of the extent of the coronal line region in COS168. We measure \NeV$\lambda$3427 emission from COS168 and provide RM time lag measurements for the emission line. Our findings are as follows:

\begin{enumerate}
    \item We measure the fractional variability of \NeVfull\ in COS168 to be $\sim$ 8\% with an SNR2 $\approx$ 29. Given the apparent variability in the \NeVfull\ light curve, this suggests that \NeVfull\ is responding to continuum variability on shorter timescales and \NeVfull\ may be emitted from a region closer than the less variable NLR in COS168.

    \item We measure the extent of the coronal line region to be \NeVRestLag\ light days. We find the coronal line region is orders of magnitude closer than some previous estimates \citep{Prieto2005, Mazzalay2013, Negus2021} in agreement with the measured size of the \CLR\ from \cite{Gravity2021}. However, our measurement (and \citealp{GRAVITY2019}) is sensitive only to the size of the inner coronal line region, while other experiments were sensitive only to the maximal size for \NeVfull\ emission. Thus, it is likely that there is a broad range of scales for \NeVfull\ emission, though the innermost regions of AGN usually contain the majority of \NeVfull\ emission.
    \item We find that the coronal line region is further extended from the center of the AGN than the BLR measured from \Ha\ and \Hb\ (see Figure \ref{fig:cartoonfig}) and the dust sublimation radius.
    \item We find that the virial product measured using coronal line emission in the RMS and median spectra is comparable to the virial product measured using the more conventional broad line emission. We also find that the virial product measured using the line width and FWHM of \Ha\ and \NeV$\lambda$3427 yield black hole masses that agree with each other and with measurements from \cite{Chen2018}. These findings imply that the inner coronal line region is plausibly in a gravitationally bound orbit around the SMBH and thus virialized. Additionally, similarities in the virial products from the median flux profile \NeVfull\ and RMS \Ha\ suggest that \NeVfull\ could plausibly be used for single-epoch BH mass measurements.

\end{enumerate}

While the results of this study are interesting, we note that this study only examines one object. Future studies should investigate coronal line emission in multiple AGN to see if these findings remain true for AGN with coronal line variability. For instance, we intend to do a broad search for coronal line variability in AGN and measure coronal line lags in these AGN. In particular, we intend to investigate whether or not the \CLR\ produces similar lags to the lag measured in COS168. Additionally, we intend to examine whether or not the \CLR\ exhibits similar virial products to the BLR in other AGN besides COS168. Investigation of \NeVfull emission in a larger quasar population will reveal if the virialized coronal line region of COS168 is the rule or the exception, with important consequences for the use of coronal line emission for black hole mass estimation.

\section{Acknowledgments}

TBS acknowledges support from a UConn SURF award. LBF, JRT, MCD, and HWS acknowledge support from NSF grants CAREER-1945546 and AST-2007993.

YS acknowledges support from NSF grant AST-2009947.

FEB acknowledges support from ANID-Chile BASAL CATA FB210003, FONDECYT Regular 1241005, and Millennium Science Initiative, AIM23-0001.

JDM's research was supported by an appointment to the NASA Postdoctoral Program at the NASA Goddard Space Flight Center, administered by Oak Ridge Associated Universities under contract with NASA.

Funding for the Sloan Digital Sky Survey V has been provided by the Alfred P. Sloan Foundation, the Heising-Simons Foundation, the National Science Foundation, and the Participating Institutions. SDSS acknowledges support and resources from the Center for High-Performance Computing at the University of Utah. SDSS telescopes are located at Apache Point Observatory, funded by the Astrophysical Research Consortium and operated by New Mexico State University, and at Las Campanas Observatory, operated by the Carnegie Institution for Science. The SDSS website is \url{www.sdss.org}.

SDSS is managed by the Astrophysical Research Consortium for the Participating Institutions of the SDSS Collaboration, including Caltech, The Carnegie Institution for Science, Chilean National Time Allocation Committee (CNTAC) ratified researchers, The Flatiron Institute, the Gotham Participation Group, Harvard University, Heidelberg University, The Johns Hopkins University, L'Ecole polytechnique f\'{e}d\'{e}rale de Lausanne (EPFL), Leibniz-Institut f\"{u}r Astrophysik Potsdam (AIP), Max-Planck-Institut f\"{u}r Astronomie (MPIA Heidelberg), Max-Planck-Institut f\"{u}r Extraterrestrische Physik (MPE), Nanjing University, National Astronomical Observatories of China (NAOC), New Mexico State University, The Ohio State University, Pennsylvania State University, Smithsonian Astrophysical Observatory, Space Telescope Science Institute (STScI), the Stellar Astrophysics Participation Group, Universidad Nacional Aut\'{o}noma de M\'{e}xico, University of Arizona, University of Colorado Boulder, University of Illinois at Urbana-Champaign, University of Toronto, University of Utah, University of Virginia, Yale University, and Yunnan University.

\newpage

\bibliography{lib}{}

\begin{thebibliography}{}
\expandafter\ifx\csname natexlab\endcsname\relax\def\natexlab#1{#1}\fi
\providecommand{\url}[1]{\href{#1}{#1}}

\bibitem[{{Baskin} {et~al.}(2014){Baskin}, {Laor}, \& {Stern}}]{Baskin2014}
{Baskin}, A., {Laor}, A., \& {Stern}, J. 2014, \mnras, 438, 604

\bibitem[{{Bentz} {et~al.}(2010){Bentz}, {Walsh}, {Barth}, {Yoshii}, {Woo}, {Wang}, {Treu}, {Thornton}, {Street}, {Steele}, {Silverman}, {Serduke}, {Sakata}, {Minezaki}, {Malkan}, {Li}, {Lee}, {Hiner}, {Hidas}, {Greene}, {Gates}, {Ganeshalingam}, {Filippenko}, {Canalizo}, {Bennert}, \& {Baliber}}]{Bentz2010}
{Bentz}, M.~C., {Walsh}, J.~L., {Barth}, A.~J., {et~al.} 2010, \apj, 716, 993

\bibitem[{{Blandford} \& {McKee}(1982)}]{Blandford1982}
{Blandford}, R.~D., \& {McKee}, C.~F. 1982, \apj, 255, 419

\bibitem[{{Bolton} {et~al.}(2012){Bolton}, {Schlegel}, {Aubourg}, {Bailey}, {Bhardwaj}, {Brownstein}, {Burles}, {Chen}, {Dawson}, {Eisenstein}, {Gunn}, {Knapp}, {Loomis}, {Lupton}, {Maraston}, {Muna}, {Myers}, {Olmstead}, {Padmanabhan}, {P{\^a}ris}, {Percival}, {Petitjean}, {Rockosi}, {Ross}, {Schneider}, {Shu}, {Strauss}, {Thomas}, {Tremonti}, {Wake}, {Weaver}, \& {Wood-Vasey}}]{Bolton2012}
{Bolton}, A.~S., {Schlegel}, D.~J., {Aubourg}, {\'E}., {et~al.} 2012, \aj, 144, 144

\bibitem[{{Bundy} {et~al.}(2015){Bundy}, {Bershady}, {Law}, {Yan}, {Drory}, {MacDonald}, {Wake}, {Cherinka}, {S{\'a}nchez-Gallego}, {Weijmans}, {Thomas}, {Tremonti}, {Masters}, {Coccato}, {Diamond-Stanic}, {Arag{\'o}n-Salamanca}, {Avila-Reese}, {Badenes}, {Falc{\'o}n-Barroso}, {Belfiore}, {Bizyaev}, {Blanc}, {Bland-Hawthorn}, {Blanton}, {Brownstein}, {Byler}, {Cappellari}, {Conroy}, {Dutton}, {Emsellem}, {Etherington}, {Frinchaboy}, {Fu}, {Gunn}, {Harding}, {Johnston}, {Kauffmann}, {Kinemuchi}, {Klaene}, {Knapen}, {Leauthaud}, {Li}, {Lin}, {Maiolino}, {Malanushenko}, {Malanushenko}, {Mao}, {Maraston}, {McDermid}, {Merrifield}, {Nichol}, {Oravetz}, {Pan}, {Parejko}, {Sanchez}, {Schlegel}, {Simmons}, {Steele}, {Steinmetz}, {Thanjavur}, {Thompson}, {Tinker}, {van den Bosch}, {Westfall}, {Wilkinson}, {Wright}, {Xiao}, \& {Zhang}}]{Bundy2015}
{Bundy}, K., {Bershady}, M.~A., {Law}, D.~R., {et~al.} 2015, \apj, 798, 7

\bibitem[{{Cackett} {et~al.}(2021){Cackett}, {Bentz}, \& {Kara}}]{Cackett2021}
{Cackett}, E.~M., {Bentz}, M.~C., \& {Kara}, E. 2021, iScience, 24, 102557

\bibitem[{{Chen} {et~al.}(2018){Chen}, {Pan}, {Pang}, \& {Huang}}]{Chen2018}
{Chen}, Z.-F., {Pan}, D.-S., {Pang}, T.-T., \& {Huang}, Y. 2018, \apjs, 234, 16

\bibitem[{{Dawson} {et~al.}(2013){Dawson}, {Schlegel}, {Ahn}, {Anderson}, {Aubourg}, {Bailey}, {Barkhouser}, {Bautista}, {Beifiori}, {Berlind}, {Bhardwaj}, {Bizyaev}, {Blake}, {Blanton}, {Blomqvist}, {Bolton}, {Borde}, {Bovy}, {Brandt}, {Brewington}, {Brinkmann}, {Brown}, {Brownstein}, {Bundy}, {Busca}, {Carithers}, {Carnero}, {Carr}, {Chen}, {Comparat}, {Connolly}, {Cope}, {Croft}, {Cuesta}, {da Costa}, {Davenport}, {Delubac}, {de Putter}, {Dhital}, {Ealet}, {Ebelke}, {Eisenstein}, {Escoffier}, {Fan}, {Filiz Ak}, {Finley}, {Font-Ribera}, {G{\'e}nova-Santos}, {Gunn}, {Guo}, {Haggard}, {Hall}, {Hamilton}, {Harris}, {Harris}, {Ho}, {Hogg}, {Holder}, {Honscheid}, {Huehnerhoff}, {Jordan}, {Jordan}, {Kauffmann}, {Kazin}, {Kirkby}, {Klaene}, {Kneib}, {Le Goff}, {Lee}, {Long}, {Loomis}, {Lundgren}, {Lupton}, {Maia}, {Makler}, {Malanushenko}, {Malanushenko}, {Mandelbaum}, {Manera}, {Maraston}, {Margala}, {Masters}, {McBride}, {McDonald}, {McGreer}, {McMahon}, {Mena}, {Miralda-Escud{\'e}}, {Montero-Dorta},
  {Montesano}, {Muna}, {Myers}, {Naugle}, {Nichol}, {Noterdaeme}, {Nuza}, {Olmstead}, {Oravetz}, {Oravetz}, {Owen}, {Padmanabhan}, {Palanque-Delabrouille}, {Pan}, {Parejko}, {P{\^a}ris}, {Percival}, {P{\'e}rez-Fournon}, {P{\'e}rez-R{\`a}fols}, {Petitjean}, {Pfaffenberger}, {Pforr}, {Pieri}, {Prada}, {Price-Whelan}, {Raddick}, {Rebolo}, {Rich}, {Richards}, {Rockosi}, {Roe}, {Ross}, {Ross}, {Rossi}, {Rubi{\~n}o-Martin}, {Samushia}, {S{\'a}nchez}, {Sayres}, {Schmidt}, {Schneider}, {Sc{\'o}ccola}, {Seo}, {Shelden}, {Sheldon}, {Shen}, {Shu}, {Slosar}, {Smee}, {Snedden}, {Stauffer}, {Steele}, {Strauss}, {Streblyanska}, {Suzuki}, {Swanson}, {Tal}, {Tanaka}, {Thomas}, {Tinker}, {Tojeiro}, {Tremonti}, {Vargas Maga{\~n}a}, {Verde}, {Viel}, {Wake}, {Watson}, {Weaver}, {Weinberg}, {Weiner}, {West}, {White}, {Wood-Vasey}, {Yeche}, {Zehavi}, {Zhao}, \& {Zheng}}]{Dawson2013}
{Dawson}, K.~S., {Schlegel}, D.~J., {Ahn}, C.~P., {et~al.} 2013, \aj, 145, 10

\bibitem[{{Doan} {et~al.}(2025){Doan}, {Satyapal}, {Reefe}, {Sexton}, {Matzko}, {McKaig}, {Secrest}, {Cann}, {Laor}, \& {Canalizo}}]{Doan2025}
{Doan}, S., {Satyapal}, S., {Reefe}, M., {et~al.} 2025, arXiv e-prints, arXiv:2501.17067

\bibitem[{{Donnan} {et~al.}(2021){Donnan}, {Horne}, \& {Hern{\'a}ndez Santisteban}}]{Fergus2021}
{Donnan}, F.~R., {Horne}, K., \& {Hern{\'a}ndez Santisteban}, J.~V. 2021, \mnras, 508, 5449

\bibitem[{{Du} {et~al.}(2014){Du}, {Hu}, {Lu}, {Wang}, {Qiu}, {Li}, {Bai}, {Kaspi}, {Netzer}, {Wang}, \& {SEAMBH Collaboration}}]{Du2014}
{Du}, P., {Hu}, C., {Lu}, K.-X., {et~al.} 2014, \apj, 782, 45

\bibitem[{{Earl} {et~al.}(2022){Earl}, {Tollerud}, {Jones}, {O'Steen}, {Kerzendorf}, {Busko}, {Shaileshahuja}, {D'Avella}, {Robitaille}, {Ginsburg}, {Homeier}, {Sip{\H{o}}cz}, {Averbukh}, {Tocknell}, {Cherinka}, {Ogaz}, {Geda}, {Lim}, {Davies}, {G{\"u}nther}, {Barbary}, {Foster}, {Conroy}, {Droettboom}, {Torres}, {Bray}, {Casey}, {Teuben}, {Crawford}, \& {Ferguson}}]{Earl2022}
{Earl}, N., {Tollerud}, E., {Jones}, C., {et~al.} 2022, {astropy/specutils: V1.7.0}, vv1.7.0,  Zenodo, doi:10.5281/zenodo.6207491

\bibitem[{{Erkens} {et~al.}(1997){Erkens}, {Appenzeller}, \& {Wagner}}]{Erkens1997}
{Erkens}, U., {Appenzeller}, I., \& {Wagner}, S. 1997, \aap, 323, 707

\bibitem[{{Event Horizon Telescope Collaboration} {et~al.}(2019){Event Horizon Telescope Collaboration}, {Akiyama}, {Alberdi}, {Alef}, {Asada}, {Azulay}, {Baczko}, {Ball}, {Balokovi{\'c}}, {Barrett}, {Bintley}, {Blackburn}, {Boland}, {Bouman}, {Bower}, {Bremer}, {Brinkerink}, {Brissenden}, {Britzen}, {Broderick}, {Broguiere}, {Bronzwaer}, {Byun}, {Carlstrom}, {Chael}, {Chan}, {Chatterjee}, {Chatterjee}, {Chen}, {Chen}, {Cho}, {Christian}, {Conway}, {Cordes}, {Crew}, {Cui}, {Davelaar}, {De Laurentis}, {Deane}, {Dempsey}, {Desvignes}, {Dexter}, {Doeleman}, {Eatough}, {Falcke}, {Fish}, {Fomalont}, {Fraga-Encinas}, {Freeman}, {Friberg}, {Fromm}, {G{\'o}mez}, {Galison}, {Gammie}, {Garc{\'\i}a}, {Gentaz}, {Georgiev}, {Goddi}, {Gold}, {Gu}, {Gurwell}, {Hada}, {Hecht}, {Hesper}, {Ho}, {Ho}, {Honma}, {Huang}, {Huang}, {Hughes}, {Ikeda}, {Inoue}, {Issaoun}, {James}, {Jannuzi}, {Janssen}, {Jeter}, {Jiang}, {Johnson}, {Jorstad}, {Jung}, {Karami}, {Karuppusamy}, {Kawashima}, {Keating}, {Kettenis}, {Kim}, {Kim}, {Kim},
  {Kino}, {Koay}, {Koch}, {Koyama}, {Kramer}, {Kramer}, {Krichbaum}, {Kuo}, {Lauer}, {Lee}, {Li}, {Li}, {Lindqvist}, {Liu}, {Liuzzo}, {Lo}, {Lobanov}, {Loinard}, {Lonsdale}, {Lu}, {MacDonald}, {Mao}, {Markoff}, {Marrone}, {Marscher}, {Mart{\'\i}-Vidal}, {Matsushita}, {Matthews}, {Medeiros}, {Menten}, {Mizuno}, {Mizuno}, {Moran}, {Moriyama}, {Moscibrodzka}, {M{\"u}ller}, {Nagai}, {Nagar}, {Nakamura}, {Narayan}, {Narayanan}, {Natarajan}, {Neri}, {Ni}, {Noutsos}, {Okino}, {Olivares}, {Ortiz-Le{\'o}n}, {Oyama}, {{\"O}zel}, {Palumbo}, {Patel}, {Pen}, {Pesce}, {Pi{\'e}tu}, {Plambeck}, {PopStefanija}, {Porth}, {Prather}, {Preciado-L{\'o}pez}, {Psaltis}, {Pu}, {Ramakrishnan}, {Rao}, {Rawlings}, {Raymond}, {Rezzolla}, {Ripperda}, {Roelofs}, {Rogers}, {Ros}, {Rose}, {Roshanineshat}, {Rottmann}, {Roy}, {Ruszczyk}, {Ryan}, {Rygl}, {S{\'a}nchez}, {S{\'a}nchez-Arguelles}, {Sasada}, {Savolainen}, {Schloerb}, {Schuster}, {Shao}, {Shen}, {Small}, {Sohn}, {SooHoo}, {Tazaki}, {Tiede}, {Tilanus}, {Titus}, {Toma}, {Torne},
  {Trent}, {Trippe}, {Tsuda}, {van Bemmel}, {van Langevelde}, {van Rossum}, {Wagner}, {Wardle}, {Weintroub}, {Wex}, {Wharton}, {Wielgus}, {Wong}, {Wu}, {Young}, {Young}, {Younsi}, {Yuan}, {Yuan}, {Zensus}, {Zhao}, {Zhao}, {Zhu}, {Algaba}, {Allardi}, {Amestica}, {Anczarski}, {Bach}, {Baganoff}, {Beaudoin}, {Benson}, {Berthold}, {Blanchard}, {Blundell}, {Bustamente}, {Cappallo}, {Castillo-Dom{\'\i}nguez}, {Chang}, {Chang}, {Chang}, {Chen}, {Chilson}, {Chuter}, {C{\'o}rdova Rosado}, {Coulson}, {Crawford}, {Crowley}, {David}, {Derome}, {Dexter}, {Dornbusch}, {Dudevoir}, {Dzib}, {Eckart}, {Eckert}, {Erickson}, {Everett}, {Faber}, {Farah}, {Fath}, {Folkers}, {Forbes}, {Freund}, {G{\'o}mez-Ruiz}, {Gale}, {Gao}, {Geertsema}, {Graham}, {Greer}, {Grosslein}, {Gueth}, {Haggard}, {Halverson}, {Han}, {Han}, {Hao}, {Hasegawa}, {Henning}, {Hern{\'a}ndez-G{\'o}mez}, {Herrero-Illana}, {Heyminck}, {Hirota}, {Hoge}, {Huang}, {Impellizzeri}, {Jiang}, {Kamble}, {Keisler}, {Kimura}, {Kono}, {Kubo}, {Kuroda}, {Lacasse}, {Laing},
  {Leitch}, {Li}, {Lin}, {Liu}, {Liu}, {Lu}, {Marson}, {Martin-Cocher}, {Massingill}, {Matulonis}, {McColl}, {McWhirter}, {Messias}, {Meyer-Zhao}, {Michalik}, {Monta{\~n}a}, {Montgomerie}, {Mora-Klein}, {Muders}, {Nadolski}, {Navarro}, {Neilsen}, {Nguyen}, {Nishioka}, {Norton}, {Nowak}, {Nystrom}, {Ogawa}, {Oshiro}, {Oyama}, {Parsons}, {Paine}, {Pe{\~n}alver}, {Phillips}, {Poirier}, {Pradel}, {Primiani}, {Raffin}, {Rahlin}, {Reiland}, {Risacher}, {Ruiz}, {S{\'a}ez-Mada{\'\i}n}, {Sassella}, {Schellart}, {Shaw}, {Silva}, {Shiokawa}, {Smith}, {Snow}, {Souccar}, {Sousa}, {Sridharan}, {Srinivasan}, {Stahm}, {Stark}, {Story}, {Timmer}, {Vertatschitsch}, {Walther}, {Wei}, {Whitehorn}, {Whitney}, {Woody}, {Wouterloot}, {Wright}, {Yamaguchi}, {Yu}, {Zeballos}, {Zhang}, \& {Ziurys}}]{EHT2019}
{Event Horizon Telescope Collaboration}, {Akiyama}, K., {Alberdi}, A., {et~al.} 2019, \apjl, 875, L1

\bibitem[{{Foltz} {et~al.}(1981){Foltz}, {Peterson}, {Cariotti}, {Byard}, {Bertram}, \& {Lawrie}}]{Foltz1981}
{Foltz}, C.~B., {Peterson}, B.~M., {Cariotti}, E.~R., {et~al.} 1981, \apj, 250, 508

\bibitem[{{Fries} {et~al.}(2023){Fries}, {Trump}, {Davis}, {Grier}, {Shen}, {Anderson}, {Dwelly}, {Eracleous}, {Homayouni}, {Horne}, {Krumpe}, {Morrison}, {Runnoe}, {Trakhtenbrot}, {Assef}, {Brandt}, {Brownstein}, {Dabbieri}, {Fix}, {Fonseca Alvarez}, {Frederick}, {Hall}, {Koekemoer}, {Li}, {Liu}, {Mart{\'\i}nez-Aldama}, {Ricci}, {Schneider}, {Sharp}, {Temple}, {Yang}, {Zeltyn}, \& {Bizyaev}}]{Fries2023}
{Fries}, L.~B., {Trump}, J.~R., {Davis}, M.~C., {et~al.} 2023, \apj, 948, 5

\bibitem[{{Gelbord} {et~al.}(2009){Gelbord}, {Mullaney}, \& {Ward}}]{Gelbord2009}
{Gelbord}, J.~M., {Mullaney}, J.~R., \& {Ward}, M.~J. 2009, \mnras, 397, 172

\bibitem[{{Gravity Collaboration} {et~al.}(2018){Gravity Collaboration}, {Sturm}, {Dexter}, {Pfuhl}, {Stock}, {Davies}, {Lutz}, {Cl{\'e}net}, {Eckart}, {Eisenhauer}, {Genzel}, {Gratadour}, {H{\"o}nig}, {Kishimoto}, {Lacour}, {Millour}, {Netzer}, {Perrin}, {Peterson}, {Petrucci}, {Rouan}, {Waisberg}, {Woillez}, {Amorim}, {Brandner}, {F{\"o}rster Schreiber}, {Garcia}, {Gillessen}, {Ott}, {Paumard}, {Perraut}, {Scheithauer}, {Straubmeier}, {Tacconi}, \& {Widmann}}]{Sturm2018}
{Gravity Collaboration}, {Sturm}, E., {Dexter}, J., {et~al.} 2018, \nat, 563, 657

\bibitem[{{GRAVITY Collaboration} {et~al.}(2019){GRAVITY Collaboration}, {Abuter}, {Amorim}, {Baub{\"o}ck}, {Berger}, {Bonnet}, {Brandner}, {Cl{\'e}net}, {Coud{\'e} Du Foresto}, {de Zeeuw}, {Dexter}, {Duvert}, {Eckart}, {Eisenhauer}, {F{\"o}rster Schreiber}, {Garcia}, {Gao}, {Gendron}, {Genzel}, {Gerhard}, {Gillessen}, {Habibi}, {Haubois}, {Henning}, {Hippler}, {Horrobin}, {Jim{\'e}nez-Rosales}, {Jocou}, {Kervella}, {Lacour}, {Lapeyr{\`e}re}, {Le Bouquin}, {L{\'e}na}, {Ott}, {Paumard}, {Perraut}, {Perrin}, {Pfuhl}, {Rabien}, {Rodriguez Coira}, {Rousset}, {Scheithauer}, {Sternberg}, {Straub}, {Straubmeier}, {Sturm}, {Tacconi}, {Vincent}, {von Fellenberg}, {Waisberg}, {Widmann}, {Wieprecht}, {Wiezorrek}, {Woillez}, \& {Yazici}}]{GRAVITY2019}
{GRAVITY Collaboration}, {Abuter}, R., {Amorim}, A., {et~al.} 2019, \aap, 625, L10

\bibitem[{{GRAVITY Collaboration} {et~al.}(2021){GRAVITY Collaboration}, {Amorim}, {Baub{\"o}ck}, {Brandner}, {Bolzer}, {Cl{\'e}net}, {Davies}, {de Zeeuw}, {Dexter}, {Drescher}, {Eckart}, {Eisenhauer}, {F{\"o}rster Schreiber}, {Gao}, {Garcia}, {Genzel}, {Gillessen}, {Gratadour}, {H{\"o}nig}, {Kaltenbrunner}, {Kishimoto}, {Lacour}, {Lutz}, {Millour}, {Netzer}, {Ott}, {Paumard}, {Perraut}, {Perrin}, {Peterson}, {Petrucci}, {Pfuhl}, {Prieto}, {Rouan}, {Sanchez-Bermudez}, {Shangguan}, {Shimizu}, {Schartmann}, {Stadler}, {Sternberg}, {Straub}, {Straubmeier}, {Sturm}, {Tacconi}, {Tristram}, {Vermot}, {von Fellenberg}, {Waisberg}, {Widmann}, \& {Woillez}}]{Gravity2021}
{GRAVITY Collaboration}, {Amorim}, A., {Baub{\"o}ck}, M., {et~al.} 2021, \aap, 648, A117

\bibitem[{{Grier} {et~al.}(2017{\natexlab{a}}){Grier}, {Pancoast}, {Barth}, {Fausnaugh}, {Brewer}, {Treu}, \& {Peterson}}]{Grier2017b}
{Grier}, C.~J., {Pancoast}, A., {Barth}, A.~J., {et~al.} 2017{\natexlab{a}}, \apj, 849, 146

\bibitem[{{Grier} {et~al.}(2013){Grier}, {Martini}, {Watson}, {Peterson}, {Bentz}, {Dasyra}, {Dietrich}, {Ferrarese}, {Pogge}, \& {Zu}}]{Grier2013}
{Grier}, C.~J., {Martini}, P., {Watson}, L.~C., {et~al.} 2013, \apj, 773, 90

\bibitem[{{Grier} {et~al.}(2017{\natexlab{b}}){Grier}, {Trump}, {Shen}, {Horne}, {Kinemuchi}, {McGreer}, {Starkey}, {Brandt}, {Hall}, {Kochanek}, {Chen}, {Denney}, {Greene}, {Ho}, {Homayouni}, {I-Hsiu Li}, {Pei}, {Peterson}, {Petitjean}, {Schneider}, {Sun}, {AlSayyad}, {Bizyaev}, {Brinkmann}, {Brownstein}, {Bundy}, {Dawson}, {Eftekharzadeh}, {Fernandez-Trincado}, {Gao}, {Hutchinson}, {Jia}, {Jiang}, {Oravetz}, {Pan}, {Paris}, {Ponder}, {Peters}, {Rogerson}, {Simmons}, {Smith}, \& {Wang}}]{Grier2017a}
{Grier}, C.~J., {Trump}, J.~R., {Shen}, Y., {et~al.} 2017{\natexlab{b}}, \apj, 851, 21

\bibitem[{{Gunn} {et~al.}(2006){Gunn}, {Siegmund}, {Mannery}, {Owen}, {Hull}, {Leger}, {Carey}, {Knapp}, {York}, {Boroski}, {Kent}, {Lupton}, {Rockosi}, {Evans}, {Waddell}, {Anderson}, {Annis}, {Barentine}, {Bartoszek}, {Bastian}, {Bracker}, {Brewington}, {Briegel}, {Brinkmann}, {Brown}, {Carr}, {Czarapata}, {Drennan}, {Dombeck}, {Federwitz}, {Gillespie}, {Gonzales}, {Hansen}, {Harvanek}, {Hayes}, {Jordan}, {Kinney}, {Klaene}, {Kleinman}, {Kron}, {Kresinski}, {Lee}, {Limmongkol}, {Lindenmeyer}, {Long}, {Loomis}, {McGehee}, {Mantsch}, {Neilsen}, {Neswold}, {Newman}, {Nitta}, {Peoples}, {Pier}, {Prieto}, {Prosapio}, {Rivetta}, {Schneider}, {Snedden}, \& {Wang}}]{Gunn2006}
{Gunn}, J.~E., {Siegmund}, W.~A., {Mannery}, E.~J., {et~al.} 2006, \aj, 131, 2332

\bibitem[{Guo {et~al.}(2020)Guo, Shen, He, Wang, Liu, Wang, Sun, Yang, Kong, \& Sheng}]{Guo2020}
Guo, H., Shen, Y., He, Z., {et~al.} 2020, The Astrophysical Journal, 888, 58.
\newblock \url{https://dx.doi.org/10.3847/1538-4357/ab5db0}

\bibitem[{{Hainline} {et~al.}(2013){Hainline}, {Hickox}, {Greene}, {Myers}, \& {Zakamska}}]{Hainline2013}
{Hainline}, K.~N., {Hickox}, R., {Greene}, J.~E., {Myers}, A.~D., \& {Zakamska}, N.~L. 2013, \apj, 774, 145

\bibitem[{{Homayouni} {et~al.}(2020){Homayouni}, {Trump}, {Grier}, {Horne}, {Shen}, {Brandt}, {Dawson}, {Alvarez}, {Green}, {Hall}, {Hern{\'a}ndez Santisteban}, {Ho}, {Kinemuchi}, {Kochanek}, {Li}, {Peterson}, {Schneider}, {Starkey}, {Bizyaev}, {Pan}, {Oravetz}, \& {Simmons}}]{Homayouni2020}
{Homayouni}, Y., {Trump}, J.~R., {Grier}, C.~J., {et~al.} 2020, \apj, 901, 55

\bibitem[{{Kollmeier} {et~al.}(2025){Kollmeier}, {Rix}, {Aerts}, {Aird}, {Alfaro}, {Almeida}, {Anderson}, {Jim{\'e}nez Arranz}, {Arseneau}, {Assef}, {Aviram}, {Aydar}, {Badenes}, {Bandyopadhyay}, {Barger}, {Barkhouser}, {Bauer}, {Bender}, {Besser}, {Bhattarai}, {Bilgi}, {Bird}, {Bizyaev}, {Blanc}, {Blanton}, {Bochanski}, {Bovy}, {Brandon}, {Brandt}, {Brownstein}, {Buchner}, {Burchett}, {Carlberg}, {Casey}, {Castaneda-Carlos}, {Chakraborty}, {Chanam{\'e}}, {Chandra}, {Cherinka}, {Chilingarian}, {Comparat}, {Cosens}, {Covey}, {Crane}, {Crumpler}, {Cunha}, {Cunningham}, {Dai}, {Darling}, {Davidson}, {Davis}, {De Lee}, {Deacon}, {M{\'e}ndez Delgado}, {Demasi}, {Demianenko}, {Derwent}, {D'Onghia}, {Di Mille}, {Dias}, {Donor}, {Drory}, {Dwelly}, {Egorov}, {Egorova}, {El-Badry}, {Engelman}, {Eracleous}, {Fan}, {Farr}, {Fries}, {Frinchaboy}, {Froning}, {G{\"a}nsicke}, {Garc{\'\i}a}, {Gelfand}, {Gentile Fusillo}, {Glover}, {Grabowski}, {Grebel}, {Green}, {Grier}, {Gupta}, {Gray}, {H{\"a}berle}, {Hall}, {Hammond},
  {Hawkins}, {Harding}, {Heged{\H{u}}s}, {Herbst}, {Hermes}, {Rodr{\'\i}guez Hidalgo}, {Hilder}, {Hogg}, {Holtzman}, {Horta}, {Huang}, {Hwang}, {Ibarra-Medel}, {Imig}, {Inight}, {Jana}, {Ji}, {Jofre}, {Johns}, {Johnson}, {Johnson}, {Johnston}, {Jones}, {Katkov}, {Koekemoer}, {Kounkel}, {Kreckel}, {Krishnarao}, {Krumpe}, {Kumari}, {Kupfer}, {Lacerna}, {Laporte}, {Lepine}, {Li}, {Liu}, {Loebman}, {Long}, {Roman-Lopes}, {Lu}, {Majewski}, {Maoz}, {McKinnon}, {Medan}, {Merloni}, {Minniti}, {Morrison}, {Myers}, {M{\'e}sz{\'a}ros}, {Nandra}, {Nayak}, {Ness}, {Nidever}, {O'Brien}, {Oeur}, {Oravetz}, {Oravetz}, {Otto}, {Adamane Pallathadka}, {Palunas}, {Pan}, {Pappalardo}, {Pandey}, {Negrete Pe{\~n}aloza}, {Pinsonneault}, {Pogge}, {Taghizadeh Popp}, {Price-Whelan}, {Pulatova}, {Qiu}, {Ramirez}, {Rankine}, {Ricci}, {Runnoe}, {Sanchez}, {Salvato}, {Sattler}, {Saydjari}, {Sayres}, {Schlaufman}, {Schneider}, {Schreiber}, {Schwope}, {Serna}, {Shen}, {Sif{\'o}n}, {Singh}, {Sinha}, {Smee}, {Song}, {Souto}, {Stassun},
  {Steinmetz}, {Stone-Martinez}, {Stringfellow}, {Stutz}, {Jos{\'e}}, {S{\'a}}, {nchez-Gallego}, {Tan}, {Tayar}, {Thai}, {Thakar}, {Ting}, {Tkachenko}, {Tovmasian}, {Trakhtenbrot}, {Fern{\'a}ndez-Trincado}, {Troup}, {Trump}, {Tuttle}, {van der Marel}, \& {Villanova}}]{Kollmeier2025}
{Kollmeier}, J.~A., {Rix}, H.-W., {Aerts}, C., {et~al.} 2025, arXiv e-prints, arXiv:2507.06989

\bibitem[{{Kormendy} \& {Ho}(2013)}]{Kormendy2013}
{Kormendy}, J., \& {Ho}, L.~C. 2013, \araa, 51, 511

\bibitem[{{Kynoch} {et~al.}(2022){Kynoch}, {Landt}, {Dehghanian}, {Ward}, \& {Ferland}}]{Kynoch2022}
{Kynoch}, D., {Landt}, H., {Dehghanian}, M., {Ward}, M.~J., \& {Ferland}, G.~J. 2022, \mnras, 516, 4397

\bibitem[{{Lamperti} {et~al.}(2017){Lamperti}, {Koss}, {Trakhtenbrot}, {Schawinski}, {Ricci}, {Oh}, {Landt}, {Riffel}, {Rodr{\'\i}guez-Ardila}, {Gehrels}, {Harrison}, {Masetti}, {Mushotzky}, {Treister}, {Ueda}, \& {Veilleux}}]{Lamperti2017}
{Lamperti}, I., {Koss}, M., {Trakhtenbrot}, B., {et~al.} 2017, \mnras, 467, 540

\bibitem[{{Landt} {et~al.}(2015{\natexlab{a}}){Landt}, {Ward}, {Steenbrugge}, \& {Ferland}}]{Landt2015b}
{Landt}, H., {Ward}, M.~J., {Steenbrugge}, K.~C., \& {Ferland}, G.~J. 2015{\natexlab{a}}, \mnras, 454, 3688

\bibitem[{{Landt} {et~al.}(2015{\natexlab{b}}){Landt}, {Ward}, {Steenbrugge}, \& {Ferland}}]{Landt2015a}
---. 2015{\natexlab{b}}, \mnras, 449, 3795

\bibitem[{{Lanzuisi} {et~al.}(2015){Lanzuisi}, {Perna}, {Delvecchio}, {Berta}, {Brusa}, {Cappelluti}, {Comastri}, {Gilli}, {Gruppioni}, {Mignoli}, {Pozzi}, {Vietri}, {Vignali}, \& {Zamorani}}]{Lanzuisi2015}
{Lanzuisi}, G., {Perna}, M., {Delvecchio}, I., {et~al.} 2015, \aap, 578, A120

\bibitem[{{Laor} \& {Draine}(1993)}]{Laor1993}
{Laor}, A., \& {Draine}, B.~T. 1993, \apj, 402, 441

\bibitem[{{Lyu} {et~al.}(2019){Lyu}, {Rieke}, \& {Smith}}]{Lyu2019}
{Lyu}, J., {Rieke}, G.~H., \& {Smith}, P.~S. 2019, \apj, 886, 33

\bibitem[{{Magorrian} {et~al.}(1998){Magorrian}, {Tremaine}, {Richstone}, {Bender}, {Bower}, {Dressler}, {Faber}, {Gebhardt}, {Green}, {Grillmair}, {Kormendy}, \& {Lauer}}]{Magorrian1998}
{Magorrian}, J., {Tremaine}, S., {Richstone}, D., {et~al.} 1998, \aj, 115, 2285

\bibitem[{{Masci} {et~al.}(2019){Masci}, {Laher}, {Rusholme}, {Shupe}, {Groom}, {Surace}, {Jackson}, {Monkewitz}, {Beck}, {Flynn}, {Terek}, {Landry}, {Hacopians}, {Desai}, {Howell}, {Brooke}, {Imel}, {Wachter}, {Ye}, {Lin}, {Cenko}, {Cunningham}, {Rebbapragada}, {Bue}, {Miller}, {Mahabal}, {Bellm}, {Patterson}, {Juri{\'c}}, {Golkhou}, {Ofek}, {Walters}, {Graham}, {Kasliwal}, {Dekany}, {Kupfer}, {Burdge}, {Cannella}, {Barlow}, {Van Sistine}, {Giomi}, {Fremling}, {Blagorodnova}, {Levitan}, {Riddle}, {Smith}, {Helou}, {Prince}, \& {Kulkarni}}]{Masci2019}
{Masci}, F.~J., {Laher}, R.~R., {Rusholme}, B., {et~al.} 2019, \pasp, 131, 018003

\bibitem[{{Matzko} {et~al.}(2025){Matzko}, {Satyapal}, {Reefe}, {McKaig}, {Sexton}, \& {Doan}}]{Matzko2025}
{Matzko}, W., {Satyapal}, S., {Reefe}, M., {et~al.} 2025, \apj, 984, 170

\bibitem[{{Mazzalay} {et~al.}(2010){Mazzalay}, {Rodr{\'\i}guez-Ardila}, \& {Komossa}}]{Mazzalay2010}
{Mazzalay}, X., {Rodr{\'\i}guez-Ardila}, A., \& {Komossa}, S. 2010, \mnras, 405, 1315

\bibitem[{{Mazzalay} {et~al.}(2013){Mazzalay}, {Rodr{\'\i}guez-Ardila}, {Komossa}, \& {McGregor}}]{Mazzalay2013}
{Mazzalay}, X., {Rodr{\'\i}guez-Ardila}, A., {Komossa}, S., \& {McGregor}, P.~J. 2013, \mnras, 430, 2411

\bibitem[{{McKaig} {et~al.}(2024){McKaig}, {Satyapal}, {Laor}, {Abel}, {Doan}, {Ricci}, \& {Cann}}]{McKaig2024}
{McKaig}, J.~D., {Satyapal}, S., {Laor}, A., {et~al.} 2024, \apj, 976, 130

\bibitem[{{Molina} {et~al.}(2021){Molina}, {Reines}, {Latimer}, {Baldassare}, \& {Salehirad}}]{Molina2021}
{Molina}, M., {Reines}, A.~E., {Latimer}, L.~J., {Baldassare}, V., \& {Salehirad}, S. 2021, \apj, 922, 155

\bibitem[{{M{\"u}ller-S{\'a}nchez} {et~al.}(2018){M{\"u}ller-S{\'a}nchez}, {Hicks}, {Malkan}, {Davies}, {Yu}, {Shaver}, \& {Davis}}]{Muller-Sanchez2018}
{M{\"u}ller-S{\'a}nchez}, F., {Hicks}, E.~K.~S., {Malkan}, M., {et~al.} 2018, \apj, 858, 48

\bibitem[{{M{\"u}ller-S{\'a}nchez} {et~al.}(2011){M{\"u}ller-S{\'a}nchez}, {Prieto}, {Hicks}, {Vives-Arias}, {Davies}, {Malkan}, {Tacconi}, \& {Genzel}}]{Muller-Sanchez2011}
{M{\"u}ller-S{\'a}nchez}, F., {Prieto}, M.~A., {Hicks}, E.~K.~S., {et~al.} 2011, \apj, 739, 69

\bibitem[{{Negus} {et~al.}(2021){Negus}, {Comerford}, {M{\"u}ller S{\'a}nchez}, {Barrera-Ballesteros}, {Drory}, {Rembold}, \& {Riffel}}]{Negus2021}
{Negus}, J., {Comerford}, J.~M., {M{\"u}ller S{\'a}nchez}, F., {et~al.} 2021, \apj, 920, 62

\bibitem[{{Negus} {et~al.}(2023){Negus}, {Comerford}, {S{\'a}nchez}, {Revalski}, {Riffel}, {Bundy}, {Nevin}, \& {Rembold}}]{Negus2023}
{Negus}, J., {Comerford}, J.~M., {S{\'a}nchez}, F.~M., {et~al.} 2023, \apj, 945, 127

\bibitem[{{Osterbrock}(1989)}]{Osterbrock1989}
{Osterbrock}, D.~E. 1989, {Astrophysics of gaseous nebulae and active galactic nuclei}

\bibitem[{{Pancoast} {et~al.}(2014){Pancoast}, {Brewer}, {Treu}, {Park}, {Barth}, {Bentz}, \& {Woo}}]{Pancoast2014}
{Pancoast}, A., {Brewer}, B.~J., {Treu}, T., {et~al.} 2014, \mnras, 445, 3073

\bibitem[{{Peterson}(1993)}]{Peterson1993}
{Peterson}, B.~M. 1993, \pasp, 105, 247

\bibitem[{{Peterson} {et~al.}(1982){Peterson}, {Foltz}, {Byard}, \& {Wagner}}]{Peterson1982}
{Peterson}, B.~M., {Foltz}, C.~B., {Byard}, P.~L., \& {Wagner}, R.~M. 1982, \apjs, 49, 469

\bibitem[{{Peterson} {et~al.}(2004){Peterson}, {Ferrarese}, {Gilbert}, {Kaspi}, {Malkan}, {Maoz}, {Merritt}, {Netzer}, {Onken}, {Pogge}, {Vestergaard}, \& {Wandel}}]{Peterson2004}
{Peterson}, B.~M., {Ferrarese}, L., {Gilbert}, K.~M., {et~al.} 2004, \apj, 613, 682

\bibitem[{{Peterson} {et~al.}(2013){Peterson}, {Denney}, {De Rosa}, {Grier}, {Pogge}, {Bentz}, {Kochanek}, {Vestergaard}, {Kilerci-Eser}, {Dalla Bont{\`a}}, \& {Ciroi}}]{Peterson2013}
{Peterson}, B.~M., {Denney}, K.~D., {De Rosa}, G., {et~al.} 2013, \apj, 779, 109

\bibitem[{{Pier} \& {Voit}(1995)}]{Pier1995}
{Pier}, E.~A., \& {Voit}, G.~M. 1995, \apj, 450, 628

\bibitem[{{Prieto} {et~al.}(2005){Prieto}, {Marco}, \& {Gallimore}}]{Prieto2005}
{Prieto}, M.~A., {Marco}, O., \& {Gallimore}, J. 2005, \mnras, 364, L28

\bibitem[{{Reefe} {et~al.}(2022){Reefe}, {Satyapal}, {Sexton}, {Doan}, {Secrest}, \& {Cann}}]{Reefe2022}
{Reefe}, M., {Satyapal}, S., {Sexton}, R.~O., {et~al.} 2022, \apj, 936, 140

\bibitem[{{Reefe} {et~al.}(2023){Reefe}, {Sexton}, {Doan}, {Satyapal}, {Secrest}, \& {Cann}}]{Reefe2023}
{Reefe}, M., {Sexton}, R.~O., {Doan}, S.~M., {et~al.} 2023, \apjs, 265, 21

\bibitem[{{Riffel} {et~al.}(2006){Riffel}, {Rodr{\'\i}guez-Ardila}, \& {Pastoriza}}]{Riffel2006}
{Riffel}, R., {Rodr{\'\i}guez-Ardila}, A., \& {Pastoriza}, M.~G. 2006, \aap, 457, 61

\bibitem[{{Rodr{\'\i}guez-Ardila} {et~al.}(2011){Rodr{\'\i}guez-Ardila}, {Prieto}, {Portilla}, \& {Tejeiro}}]{Rodriguez2011}
{Rodr{\'\i}guez-Ardila}, A., {Prieto}, M.~A., {Portilla}, J.~G., \& {Tejeiro}, J.~M. 2011, \apj, 743, 100

\bibitem[{{Rodr{\'\i}guez-Ardila} {et~al.}(2006){Rodr{\'\i}guez-Ardila}, {Prieto}, {Viegas}, \& {Gruenwald}}]{Rodriguez2006}
{Rodr{\'\i}guez-Ardila}, A., {Prieto}, M.~A., {Viegas}, S., \& {Gruenwald}, R. 2006, \apj, 653, 1098

\bibitem[{{Runnoe} {et~al.}(2012){Runnoe}, {Brotherton}, \& {Shang}}]{Runnoe2012}
{Runnoe}, J.~C., {Brotherton}, M.~S., \& {Shang}, Z. 2012, \mnras, 426, 2677

\bibitem[{{Satyapal} {et~al.}(2023){Satyapal}, {Reefe}, {Doan}, {Sexton}, {Secrest}, {Cann}, \& {Matzko}}]{Satyapal2023}
{Satyapal}, S., {Reefe}, M., {Doan}, S., {et~al.} 2023, in American Astronomical Society Meeting Abstracts, Vol. 241, American Astronomical Society Meeting Abstracts, 242.03

\bibitem[{{Sayres} {et~al.}(2022){Sayres}, {S{\'a}nchez-Gallego}, {Blanton}, {Engelman}, {Finkbeiner}, {Hogg}, {Holtzman}, {Jurgenson}, {Pogge}, {Ram{\'\i}rez}, {Saydjari}, {Schlafly}, \& {Tuttle}}]{Sayres2022}
{Sayres}, C., {S{\'a}nchez-Gallego}, J.~R., {Blanton}, M.~R., {et~al.} 2022, in Society of Photo-Optical Instrumentation Engineers (SPIE) Conference Series, Vol. 12184, Ground-based and Airborne Instrumentation for Astronomy IX, ed. C.~J. {Evans}, J.~J. {Bryant}, \& K.~{Motohara}, 121847K

\bibitem[{{Seyfert}(1943)}]{Seyfert1943}
{Seyfert}, C.~K. 1943, \apj, 97, 28

\bibitem[{{Shen} {et~al.}(2015){Shen}, {Brandt}, {Dawson}, {Hall}, {McGreer}, {Anderson}, {Chen}, {Denney}, {Eftekharzadeh}, {Fan}, {Gao}, {Green}, {Greene}, {Ho}, {Horne}, {Jiang}, {Kelly}, {Kinemuchi}, {Kochanek}, {P{\^a}ris}, {Peters}, {Peterson}, {Petitjean}, {Ponder}, {Richards}, {Schneider}, {Seth}, {Smith}, {Strauss}, {Tao}, {Trump}, {Wood-Vasey}, {Zu}, {Eisenstein}, {Pan}, {Bizyaev}, {Malanushenko}, {Malanushenko}, \& {Oravetz}}]{Shen2015}
{Shen}, Y., {Brandt}, W.~N., {Dawson}, K.~S., {et~al.} 2015, \apjs, 216, 4

\bibitem[{{Shen} {et~al.}(2019){Shen}, {Hall}, {Horne}, {Zhu}, {McGreer}, {Simm}, {Trump}, {Kinemuchi}, {Brandt}, {Green}, {Grier}, {Guo}, {Ho}, {Homayouni}, {Jiang}, {I-Hsiu Li}, {Morganson}, {Petitjean}, {Richards}, {Schneider}, {Starkey}, {Wang}, {Chambers}, {Kaiser}, {Kudritzki}, {Magnier}, \& {Waters}}]{Shen2019}
{Shen}, Y., {Hall}, P.~B., {Horne}, K., {et~al.} 2019, \apjs, 241, 34

\bibitem[{{Shen} {et~al.}(2024){Shen}, {Grier}, {Horne}, {Stone}, {Li}, {Yang}, {Homayouni}, {Trump}, {Anderson}, {Brandt}, {Hall}, {Ho}, {Jiang}, {Petitjean}, {Schneider}, {Tao}, {Donnan}, {AlSayyad}, {Bershady}, {Blanton}, {Bizyaev}, {Bundy}, {Chen}, {Davis}, {Dawson}, {Fan}, {Greene}, {Gr{\"o}ller}, {Guo}, {Ibarra-Medel}, {Jiang}, {Keenan}, {Kollmeier}, {Lejoly}, {Li}, {de la Macorra}, {Moe}, {Nie}, {Rossi}, {Smith}, {Tee}, {Weijmans}, {Xu}, {Yue}, {Zhou}, {Zhou}, \& {Zou}}]{Shen2024}
{Shen}, Y., {Grier}, C.~J., {Horne}, K., {et~al.} 2024, \apjs, 272, 26

\bibitem[{{Smee} {et~al.}(2013){Smee}, {Gunn}, {Uomoto}, {Roe}, {Schlegel}, {Rockosi}, {Carr}, {Leger}, {Dawson}, {Olmstead}, {Brinkmann}, {Owen}, {Barkhouser}, {Honscheid}, {Harding}, {Long}, {Lupton}, {Loomis}, {Anderson}, {Annis}, {Bernardi}, {Bhardwaj}, {Bizyaev}, {Bolton}, {Brewington}, {Briggs}, {Burles}, {Burns}, {Castander}, {Connolly}, {Davenport}, {Ebelke}, {Epps}, {Feldman}, {Friedman}, {Frieman}, {Heckman}, {Hull}, {Knapp}, {Lawrence}, {Loveday}, {Mannery}, {Malanushenko}, {Malanushenko}, {Merrelli}, {Muna}, {Newman}, {Nichol}, {Oravetz}, {Pan}, {Pope}, {Ricketts}, {Shelden}, {Sandford}, {Siegmund}, {Simmons}, {Smith}, {Snedden}, {Schneider}, {SubbaRao}, {Tremonti}, {Waddell}, \& {York}}]{Smee2013}
{Smee}, S.~A., {Gunn}, J.~E., {Uomoto}, A., {et~al.} 2013, \aj, 146, 32

\bibitem[{{Sun} {et~al.}(2018){Sun}, {Grier}, \& {Peterson}}]{Sun2018}
{Sun}, M., {Grier}, C.~J., \& {Peterson}, B.~M. 2018, {PyCCF: Python Cross Correlation Function for reverberation mapping studies}, Astrophysics Source Code Library, record ascl:1805.032, ,

\bibitem[{{U} {et~al.}(2022){U}, {Barth}, {Vogler}, {Guo}, {Treu}, {Bennert}, {Canalizo}, {Filippenko}, {Gates}, {Hamann}, {Joner}, {Malkan}, {Pancoast}, {Williams}, {Woo}, {Abolfathi}, {Abramson}, {Armen}, {Bae}, {Bohn}, {Boizelle}, {Bostroem}, {Brandel}, {Brink}, {Channa}, {Cooper}, {Cosens}, {Donohue}, {Fillingham}, {Gonz{\'a}lez-Buitrago}, {Halevi}, {Halle}, {Hood}, {Horne}, {Horst}, {de Kouchkovsky}, {Kuhn}, {Kumar}, {Leonard}, {Loveland}, {Manzano-King}, {McHardy}, {Michel}, {Olaes}, {Park}, {Park}, {Pei}, {Ross}, {Runco}, {Samuel}, {S{\'a}nchez}, {Scott}, {Sexton}, {Shin}, {Shivvers}, {Spencer}, {Stahl}, {Stegman}, {Stomberg}, {Valenti}, {Villafa{\~n}a}, {Walsh}, {Yuk}, \& {Zheng}}]{U2022}
{U}, V., {Barth}, A.~J., {Vogler}, H.~A., {et~al.} 2022, \apj, 925, 52

\bibitem[{{Vanden Berk} {et~al.}(2001){Vanden Berk}, {Richards}, {Bauer}, {Strauss}, {Schneider}, {Heckman}, {York}, {Hall}, {Fan}, {Knapp}, {Anderson}, {Annis}, {Bahcall}, {Bernardi}, {Briggs}, {Brinkmann}, {Brunner}, {Burles}, {Carey}, {Castander}, {Connolly}, {Crocker}, {Csabai}, {Doi}, {Finkbeiner}, {Friedman}, {Frieman}, {Fukugita}, {Gunn}, {Hennessy}, {Ivezi{\'c}}, {Kent}, {Kunszt}, {Lamb}, {Leger}, {Long}, {Loveday}, {Lupton}, {Meiksin}, {Merelli}, {Munn}, {Newberg}, {Newcomb}, {Nichol}, {Owen}, {Pier}, {Pope}, {Rockosi}, {Schlegel}, {Siegmund}, {Smee}, {Snir}, {Stoughton}, {Stubbs}, {SubbaRao}, {Szalay}, {Szokoly}, {Tremonti}, {Uomoto}, {Waddell}, {Yanny}, \& {Zheng}}]{Vanden2001}
{Vanden Berk}, D.~E., {Richards}, G.~T., {Bauer}, A., {et~al.} 2001, \aj, 122, 549

\bibitem[{{Veilleux}(1988)}]{Veilleux1988}
{Veilleux}, S. 1988, \aj, 95, 1695

\bibitem[{{Vestergaard} \& {Peterson}(2006)}]{Vestergaard2006}
{Vestergaard}, M., \& {Peterson}, B.~M. 2006, \apj, 641, 689

\bibitem[{{Woo} {et~al.}(2015){Woo}, {Yoon}, {Park}, {Park}, \& {Kim}}]{Woo2015}
{Woo}, J.-H., {Yoon}, Y., {Park}, S., {Park}, D., \& {Kim}, S.~C. 2015, \apj, 801, 38

\bibitem[{{Yin} {et~al.}(2025){Yin}, {Lawrence}, {Ward}, {Homan}, \& {Kollatschny}}]{Yin2025}
{Yin}, C., {Lawrence}, A., {Ward}, M., {Homan}, D., \& {Kollatschny}, W. 2025, \mnras, 540, 3032

\bibitem[{{York} {et~al.}(2000){York}, {Adelman}, {Anderson}, {Anderson}, {Annis}, {Bahcall}, {Bakken}, {Barkhouser}, {Bastian}, {Berman}, {Boroski}, {Bracker}, {Briegel}, {Briggs}, {Brinkmann}, {Brunner}, {Burles}, {Carey}, {Carr}, {Castander}, {Chen}, {Colestock}, {Connolly}, {Crocker}, {Csabai}, {Czarapata}, {Davis}, {Doi}, {Dombeck}, {Eisenstein}, {Ellman}, {Elms}, {Evans}, {Fan}, {Federwitz}, {Fiscelli}, {Friedman}, {Frieman}, {Fukugita}, {Gillespie}, {Gunn}, {Gurbani}, {de Haas}, {Haldeman}, {Harris}, {Hayes}, {Heckman}, {Hennessy}, {Hindsley}, {Holm}, {Holmgren}, {Huang}, {Hull}, {Husby}, {Ichikawa}, {Ichikawa}, {Ivezi{\'c}}, {Kent}, {Kim}, {Kinney}, {Klaene}, {Kleinman}, {Kleinman}, {Knapp}, {Korienek}, {Kron}, {Kunszt}, {Lamb}, {Lee}, {Leger}, {Limmongkol}, {Lindenmeyer}, {Long}, {Loomis}, {Loveday}, {Lucinio}, {Lupton}, {MacKinnon}, {Mannery}, {Mantsch}, {Margon}, {McGehee}, {McKay}, {Meiksin}, {Merelli}, {Monet}, {Munn}, {Narayanan}, {Nash}, {Neilsen}, {Neswold}, {Newberg}, {Nichol}, {Nicinski},
  {Nonino}, {Okada}, {Okamura}, {Ostriker}, {Owen}, {Pauls}, {Peoples}, {Peterson}, {Petravick}, {Pier}, {Pope}, {Pordes}, {Prosapio}, {Rechenmacher}, {Quinn}, {Richards}, {Richmond}, {Rivetta}, {Rockosi}, {Ruthmansdorfer}, {Sandford}, {Schlegel}, {Schneider}, {Sekiguchi}, {Sergey}, {Shimasaku}, {Siegmund}, {Smee}, {Smith}, {Snedden}, {Stone}, {Stoughton}, {Strauss}, {Stubbs}, {SubbaRao}, {Szalay}, {Szapudi}, {Szokoly}, {Thakar}, {Tremonti}, {Tucker}, {Uomoto}, {Vanden Berk}, {Vogeley}, {Waddell}, {Wang}, {Watanabe}, {Weinberg}, {Yanny}, {Yasuda}, \& {SDSS Collaboration}}]{York2000}
{York}, D.~G., {Adelman}, J., {Anderson}, John~E., J., {et~al.} 2000, \aj, 120, 1579

\bibitem[{{Zu} {et~al.}(2010){Zu}, {Kochanek}, \& {Peterson}}]{Zu2010}
{Zu}, Y., {Kochanek}, C.~S., \& {Peterson}, B.~M. 2010, {JAVELIN: Just Another Vehicle for Estimating Lags In Nuclei}, Astrophysics Source Code Library, record ascl:1010.007, ,

\end{thebibliography}

\end{document}